\documentclass[onecolumn,superscriptaddress]{revtex4}

\pdfoutput=1
 \usepackage{amsfonts}
\usepackage{amsmath}
\usepackage{amssymb}
\usepackage{graphicx}
\usepackage{dcolumn}
\usepackage{dsfont}
\usepackage{times}
\usepackage{epsfig}
\usepackage[]{units}

\usepackage{lipsum}

\newcommand{\R}{{\mathbb R}}

\newcommand{\C}{{\mathbb C}}

\newcommand{\pa}{\partial}

\def\epsilon{\varepsilon}

\def\q{\mathbf{q}}
\def\F{\mathbf{F}}

\def\beq{\begin{equation}}
\def\eeq{\end{equation}}
\def\l{\ell}

\def\hatd{\hat{d}}
\def\hatdelta{\hat{\delta}}

\newcommand{\subfigimg}[3][,]{%
  \setbox1=\hbox{\includegraphics[#1]{#3}}
  \leavevmode\rlap{\usebox1}
  \rlap{\hspace*{5pt}\raisebox{\dimexpr\ht1-.5\baselineskip}{#2}}
  \phantom{\usebox1}
}

\begin{document}

\title{Conical Wave Propagation and Diffraction in 2D Hexagonally Packed Granular Lattices}

\author{C. Chong}
\affiliation{Department of Mechanical and Process Engineering (D-MAVT), ETH-Zurich,  8092 Zurich, Switzerland}
\affiliation{Department of Mathematics, Bowdoin College, Brunswick, ME 04011, USA}

\author{P.G. Kevrekidis}
\affiliation{Department of Mathematics and Statistics, University of Massachusetts, Amherst, MA 01003-4515, USA}

\author{M.J. Ablowitz}
\affiliation{Department of Applied Mathematics, University of Colorado, 526 UCB, Boulder, Colorado 80309-0526, USA}

\author{Yi-Ping Ma}
\affiliation{Department of Applied Mathematics, University of Colorado, 526 UCB, Boulder, Colorado 80309-0526, USA}


\begin{abstract}
Linear and nonlinear mechanisms for conical wave propagation in
two-dimensional lattices are explored in the realm of phononic crystals.
As a prototypical example,
a statically compressed granular lattice of spherical particles arranged in a hexagonal packing configuration is analyzed. Upon identifying the dispersion relation of the
underlying linear problem, the resulting diffraction
properties are considered. Analysis both via a heuristic argument for the linear
propagation of a wavepacket, as well as via  asymptotic analysis
leading to the derivation of a Dirac system suggests the occurrence of conical diffraction.
This analysis
is valid for strong precompression i.e., near the linear regime.
For weak precompression,
conical wave propagation is still possible, but the resulting expanding circular wave front
is of a non-oscillatory nature, resulting from the complex
interplay between the discreteness, nonlinearity and geometry of the packing.
The transition between these two types of propagation is explored.

\end{abstract}

\maketitle

 \begin{figure}[t]

    \centering
  \begin{tabular}{@{}p{0.45\linewidth}@{\quad}p{0.45\linewidth}@{}}
    \subfigimg[width=\linewidth]{\bf (a)}{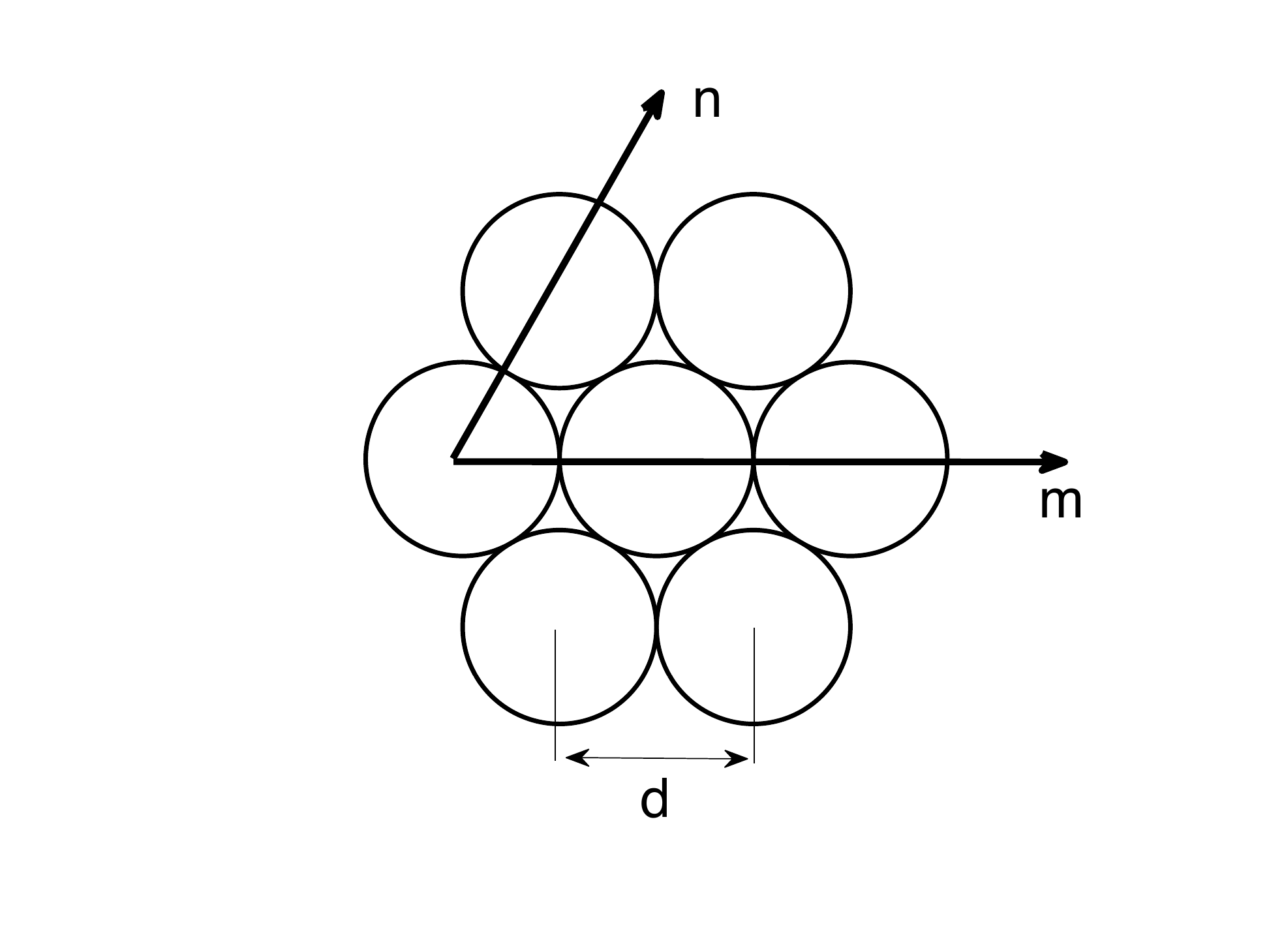} &
     \subfigimg[width=\linewidth]{\bf (b)}{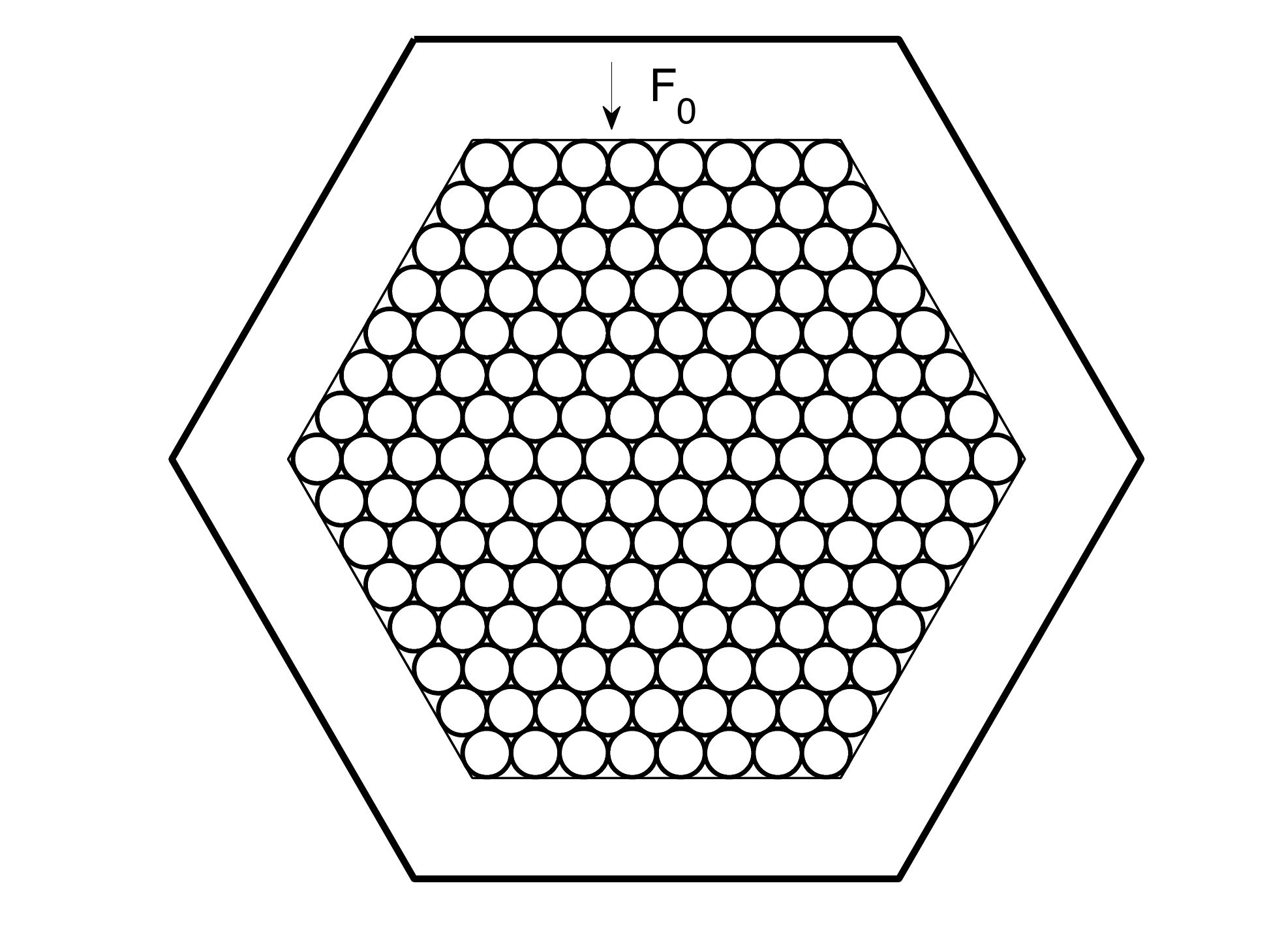}
  \end{tabular}
 \caption{\textbf{(a)} Orientation of the index convention. The $m$ axis
 represents the horizontal direction. The $m$ axis and the $n$ axis meet at an
angle of $\theta = \pi/3$. In the absence of precompression, the equilibrium distance between the centers of adjacent
 beads is the bead diameter $d$. \textbf{(b)} The hexagonal lattice is compressed uniformly on all boundaries
 which induces a static overlap $\delta$ such that the equilibrium distance between the centers of adjacent
 beads is $d - \delta$. Under these compression conditions, the static equilibrium configuration
 has the hexagonal symmetry. The amount of static force $F_0$, required to induce the static overlap will depend on the
 number of beads in contact with the boundary. The compression amount of the boundary is greatly exaggerated in this figure for clarity.}
 \label{fig:lattice}
\end{figure}

\section{Introduction}

Nearly two centuries ago, Hamilton predicted that under certain conditions, a narrow beam of light entering a crystal will spread into a hollow cone within the crystal \cite{Hamilton}.
This phenomenon, termed conical diffraction, was observed later by Lloyd \cite{Lloyd}. Conical diffraction is possible in crystals with dispersion surfaces that intersect
at a singular point where the group velocity is not uniquely defined \cite{Peleg}. This is often referred to as the Dirac point or diabolical point.
The geometry of the dispersion in its vicinity is cone-like,
and is known as a Dirac cone \cite{Hamilton}. One notable example of a physical system possessing Dirac cones that has
renewed enthusiasm in the topic is graphene. Graphene, which is a monolayer of graphite that exhibits extremely high electron mobility~\cite{Novoselov}, can be used
in a host of applications including medicine, energy, sensing and electronics~\cite{Graphene_review}.   In the case of graphene, the atoms are packed in a honeycomb structure. It is this packing
geometry that leads to Dirac points in the dispersion
relation~\cite{Graphene1,Graphene2,Graphene3}.
One important difference between Dirac points arising in honeycomb structures
and those studied by Hamilton \cite{Hamilton}, is that in the former, the Dirac points always lie at the vertices of the Brillouin zone, and hence are independent
of the specific parameters of the system, while in the latter
the singularity in $k$-space arose due to polarization.
In this sense, conical diffraction is generic in systems with
e.g. a honeycomb symmetry and indeed emerges due to the special
symmetry of the lattice.  This has led to a burst of activity towards the
study of Dirac points in other physical systems with the honeycomb and hexagonal symmetries, e.g., in photonics~\cite{HuangNature,Wang,Ochiai,Diem,Sep,Peleg,Ablowitz1,Ablowitz2,Darcy,Berry} --giving also rise to the
term ``photonic graphene''--,
where it has been shown that
conical diffraction is possible \cite{Peleg,Ablowitz1}.
More recently, Dirac points have started to be explored in phononic
systems, where pressure waves are manipulated rather than light waves \cite{Torrent,Torrent2,Craster}. The presence of Dirac points in such
phononic systems
suggests that conical diffraction is possible there too, but this possibility
has not been explored up to now, to the best of our knowledge.
Moreover, the presence of nonlinearity can play a crucial role in the dynamics,
and potentially even lead to a breakdown of the conical
wave propagation in honeycomb lattices as shown e.g. in~\cite{fail,BorisBook}.

In this present work, we investigate conical diffraction in a phononic
lattice, emphasizing the near-linear limit, but also
considering varying degrees of nonlinearity.
We chose a system that is well within the realm of ongoing experimental
considerations, namely, a two-dimensional hexagonally packed
lattice of spherical particles that interact nonlinearly through point contacts. Such systems have been termed granular crystals \cite{Nesterenko2001,review_Sen,review_Theo,review_PK},
and have been proposed for a range of applications including --but not
limited to-- shock and energy absorbing layers~\cite{Daraio2006bb,Hong2005,Fraternali2008,Doney2006}, actuating devices \cite{dev08}, acoustic lenses \cite{Spadoni}, acoustic diodes \cite{Nature11} and sound scramblers \cite{dar05,Nesterenko2005}.
Wave propagation has been studied extensively in one-dimensional (1D) granular crystals where robust highly localized waves and variants thereof
have been identified in various configurations, see the reviews {\cite{Nesterenko2001,review_Theo,review_Sen,review_PK}. Higher dimensional granular crystals
have also been studied \cite{l6,l7,Owens:2011,Kondic:2012,l10,l11,Aswasthi,ivan12,andrea,l8,Abd:2010,l9,Nishida,Coste:2008,Gilles:2003,Mouraille:2006,Leonard11}, but to a far lesser extent than in 1D.
If the particles are packed so that they are just touching, then the
resulting dynamics are purely nonlinear, and hence there is no dispersion.
However, circular patterns in such systems have been reported
on \cite{andrea}, although the exact mechanisms for their formation
have not been identified. On the other hand, if one compresses the lattice
at the boundary, a
static overlap between adjacent particles will be induced, and hence the equations become linearizable, leading to dispersion surfaces and the possibility
of Dirac points. In this paper, we study conical wave propagation in a 2D hexagonal granular lattice
as the nonlinear response is tuned from linear and weakly nonlinear
to strongly nonlinear  and the resulting transition between outward
conical diffraction and complex outward propagating
wave fronts. The paper is organized as follows. The model equation and its linearization are introduced in Sec.~\ref{sec:model}.
We show analytically in Sec.~\ref{sec:disp} that the dispersion features Dirac cones and we present a heuristic argument for conical diffraction. This
approach is corroborated by an asymptotic analysis in section IV
demonstrating the
relevance of a Dirac system for describing the dynamics in the vicinity
of the conical point.
The transition from linear to strongly nonlinear dynamics is studied numerically in Sec.~\ref{sec:numerics}.  Concluding remarks and
open problems are given in~\ref{sec:theend}.

\section{Model and linearization} \label{sec:model}

We consider a hexagonally packed lattice of spherical particles. To that end, we define the lattice basis vectors $e_1=(1,0)$ and $e_2=(1/2, \sqrt{3}/2)$.
Let $\mathbf{q}_{m,n}(t)= (x_{m,n}(t),y_{m,n}(t)) \in\R^2$ represent the displacement from the static equilibrium of the bead situated at position $\textbf{p} = d (m e_1 + n e_2)$ in the plane,
where $d$ is the bead diameter, see Fig.~\ref{fig:lattice}(a). If the lattice is precompressed
by a static force (see Fig.~\ref{fig:lattice}(b)) thereby inducing a static overlap $\delta$ between each adjacent bead (when measuring the distances between their
centers), then the modified positions of the beads in equilibrium become
\begin{equation} \label{eq:radius}
\textbf{p} = (d - \delta) (m e_1 + n e_2)
\end{equation}
Assuming deformations that are small relative to the bead diameter, the magnitude of the force resulting from elastic deformation of two spherical particles in contact is given by the classical Hertz law \cite{Hertz,Johnson},
\begin{equation} \label{eq:Hertz}
V'(r) = \gamma [ d - r ]^{3/2}_+
\end{equation}
where $r$ is the distance between the two center points of the beads and the bracket is defined by $[x]_+ = \max(0,x)$ (indicating that there is no tensile force).
$\gamma$ is a parameter depending on the elastic
properties of the material and the geometric characteristics of the
beads~\cite{Nesterenko2001}. For a uniform lattice, we have $\gamma= \frac{E \sqrt{d}}{3(1-\nu^2)}$, where $E$ is the elastic (Young's) modulus of the particle material and $\nu$
is the Poisson ratio.
By combining Eqs.~\eqref{eq:radius} and \eqref{eq:Hertz} and ignoring all other forces (plasticity, viscous damping, rotation dynamics, an approximation
that has been shown to be qualitatively reasonable in
comparison with experimental results e.g. in~\cite{andrea})
we can write the equations of
motion strictly in terms of the horizontal $x_{m,n}$ and vertical $y_{m,n}$
displacements from the equilibrium position,
\begin{equation} \label{eq:model}
\begin{array}{ll} \displaystyle
\ddot{\mathbf{q}}_{m,n} =&  \mathbf{F}_1(\mathbf{q}_{m,n} -\mathbf{q}_{m-1,n})
+ \mathbf{F}_2( \mathbf{q}_{m,n} - \mathbf{q}_{m,n-1})
- \mathbf{F}_3(\mathbf{q}_{m+1,n-1} -\mathbf{q}_{m,n})
\\
& - \mathbf{F}_1(\mathbf{q}_{m+1,n} - \mathbf{q}_{m,n})  - \mathbf{F}_2(\mathbf{q}_{m,n+1} - \mathbf{q}_{m,n})
+\mathbf{F}_3(\mathbf{q}_{m,n} - \mathbf{q}_{m-1,n+1})
\end{array}
\end{equation}
which takes into account the six contact points resulting from the hexagonal symmetry. The vector valued functions $\mathbf{F}_j(\mathbf{q}) = \mathbf{F}_j(x,y) = [ F_{j,x}(x,y), F_{j,y}(x,y) ]^T$ , $j \in \{1,2,3\}$ have the form,
$$
\begin{array}{l}
F_{1,x}(x,y) = \gamma \left[ d - \sqrt{ (d- \delta + x  )^2   + y^2}  \,    \right]^{3/2}_{+}   \frac{d - \delta + x    }{ \sqrt{ (d-\delta + x )^2   + y^2}  } \\
F_{2,x}(x,y) =  \gamma \left[ d - \sqrt{ ( (d- \delta) \cos(\theta) + x )^2   + ( (d- \delta) \sin(\theta) + y  )^2}  \,    \right]^{3/2}_{+}   \frac{ (d- \delta) \cos(\theta) + x  }{ \sqrt{ ( (d- \delta) \cos(\theta) + x )^2   + ( (d- \delta) \sin(\theta) + y  )^2} }\\
F_{3,x}(x,y) =  \gamma \left[ d - \sqrt{ ((d- \delta)\cos(\theta) + x )^2   + (  (d- \delta) \sin(-\theta) + y )^2}  \,    \right]^{3/2}_{+}   \frac{(d- \delta)\cos(\theta) + x }{ \sqrt{ ((d- \delta)\cos(\theta) + x )^2   + (  (d- \delta) \sin(-\theta) + y )^2}  } \\
F_{1,y}(x,y) =  \gamma \left[ d - \sqrt{ (d-\delta + x )^2   + y^2}  \,    \right]^{3/2}_{+}   \frac{y  }{  \sqrt{ (d-\delta + x )^2   + y^2}}   \\
F_{2,y}(x,y) =  \gamma \left[ d - \sqrt{ ((d- \delta) \cos(\theta) + x)^2   + ( (d- \delta) \sin(\theta) + y )^2}  \,    \right]^{3/2}_{+}    \frac{ (d- \delta) \sin(\theta) + y  }{ \sqrt{ ( (d- \delta) \cos(\theta) + x )^2   + ( (d- \delta) \sin(\theta) + y  )^2} } \\
F_{3,y}(x,y) =  \gamma \left[ d - \sqrt{ ((d- \delta)\cos(\theta) + x )^2   + (  (d- \delta) \sin(-\theta) + y)^2}  \,    \right]^{3/2}_{+}   \frac{(d- \delta) \sin(-\theta) + y }{ \sqrt{ ((d- \delta)\cos(\theta) + x )^2   + (  (d- \delta) \sin(-\theta) + y )^2}  }
\end{array}
$$
where  $\theta = \pi/3$.

We remark that in order for the equations of motion (\ref{eq:model}) to be valid, any bead that a given bead is in contact with must be one of its original six neighbors.
The distance (in terms of bead center) to the closest next-nearest-neighbor of any given bead at equilibrium is $\sqrt{3} (d - \delta)$. The corresponding distance between the surfaces of the (uncompressed) beads is thus $\sqrt{3} (d - \delta) - d$. This simple observation leads to
a sufficient condition to guarantee that no bead is in contact with its next-nearest-neighbor:


\begin{equation}\label{eq:suff-no-bad-contact}
\forall m,n,\quad |\mathbf{q}_{m,n}|< \frac{\sqrt{3}(d-\delta) - d}{2}.
\end{equation}
In experiments, typical displacements are small relative to the bead diameter \cite{andrea}, and thus the condition \eqref{eq:suff-no-bad-contact} would not be a concern
in such settings.

Assuming small strains, i.e. 
\begin{equation} \label{eq:smstrain}
 \frac{| \mathbf{q}_{m \pm 1,n} - \mathbf{q}_{m,n} |}{\delta} \ll 1, \quad  \frac{| \mathbf{q}_{m,n \pm 1} - \mathbf{q}_{m,n} |}{\delta} \ll 1, \quad   \frac{| \mathbf{q}_{m \pm1,n \mp 1} - \mathbf{q}_{m,n} |}{\delta} \ll 1
\end{equation}
we can make use of the Taylor expansion,
$$  \mathbf{F}_j(\mathbf{q}) \approx \mathbf{F}_j(\mathbf{q}_0) + D \mathbf{F}_j(\mathbf{q}_0)   \mathbf{q}   $$
where $D\F_j$ is the Jacobian matrix of $\F_j$. Using this notation, we write the linearized equations of motion:
\begin{equation} \label{eq:linear}
\ddot{\q}_{m,n}  =  -D \mathbf{F}_1 ( \q_{m+1,n} + \q_{m-1,n} )
- D \mathbf{F}_2 ( \q_{m,n+1} + \q_{m,n-1} )
- D \mathbf{F}_3 ( \q_{m-1,n+1} + \q_{m+1,n-1} )
+ 2( D \mathbf{F}_1 + D \mathbf{F}_2 + D \mathbf{F}_3)  \q_{m,n}
\end{equation}
where we use the following notation for the entries of the Jacobian matrices,
\begin{equation*}
DF_i = \begin{pmatrix} a_i & b_i \\ c_i & d_i\end{pmatrix}, \qquad  j \in \{1,2,3\}
\end{equation*}
with
\begin{align*}
&a_1=-\frac{3}{2}(\hatd-\hatdelta),\quad b_1=0,\quad c_1=0,\quad d_1=\hatdelta;\\
&a_2=-\frac{3}{8}(\hatd-3\hatdelta),\quad b_2=-\frac{\sqrt{3}}{8}(3\hatd-\hatdelta),\quad c_2=b_2,\quad d_2=\frac{1}{8}(11\hatdelta-9\hatd);\\
&a_3=a_2,\quad b_3=-b_2,\quad c_3=-c_2,\quad d_3=d_2.
\end{align*}
where
\begin{equation}
\hatd\equiv \frac{d\gamma\sqrt{\delta}}{d-\delta},\quad
\hatdelta\equiv \frac{\delta\gamma\sqrt{\delta}}{d-\delta}.
\end{equation}

\section{Dispersion Relation, Dirac Points and Conical Diffraction } \label{sec:disp}

Defining the 2D discrete transform,
\begin{equation}
\label{eq:DFT}
\hat{x}(k,\l) = \sum_{m,n} x_{m,n} \exp\left(i( k m + \frac{n}{2}( k +  \sqrt{3} \l  ) )\right), \qquad \hat{y}(k,\l) = \sum_{m,n} y_{m,n} \exp\left(i( k m + \frac{n}{2}( k +  \sqrt{3} \l  ) )\right)
\end{equation}
allows us to write the linear system in the frequency domain:
\begin{equation} \label{lin2}
\left.
\begin{array}{ll}
\pa_t^2 \hat{x} &= \omega_a \hat{x} + \omega_b \hat{y} \\
\pa_t^2 \hat{y} &= \omega_c \hat{x} + \omega_d \hat{y}
\end{array} \vspace{.5cm} \right\}
\end{equation}
where,
\begin{eqnarray*}
\omega_a(k,\ell) &=& -2a_1\cos(k) - 2a_2\cos(k/2+ \sqrt{3}/2 \l) - 2a_3\cos(k/2 - \sqrt{3}/2 \l) + 2(a_1+a_2+a_3) \\
\omega_b(k,\ell) &=& -2b_1\cos(k) - 2b_2\cos(k/2+ \sqrt{3}/2 \l) - 2b_3\cos(k/2 - \sqrt{3}/2 \l) + 2(b_1+b_2+b_3) \\
\omega_c(k,\ell) &=& -2c_1\cos(k) - 2c_2\cos(k/2+ \sqrt{3}/2 \l) - 2c_3\cos(k/2 - \sqrt{3}/2 \l) + 2(c_1+c_2+c_3) \\
\omega_d(k,\ell) &=& -2d_1\cos(k) - 2d_2\cos(k/2+ \sqrt{3}/2 \l) - 2d_3\cos(k/2 - \sqrt{3}/2 \l) + 2(d_1+d_2+d_3).
\end{eqnarray*}
For fixed $k$ and $\ell$ Eq.~\eqref{lin2} is solved by $v e^{- i \omega t}$, where $v\in\R^2$ and,
\begin{equation} \label{eq:disp_matrix}
-\omega^2 v = \mathcal{H}   v, \qquad  \mathcal{H}:= \begin{pmatrix}
            \omega_a & \omega_b \\
            \omega_c & \omega_d \\
           \end{pmatrix}
\end{equation}
The eigenvalues $\lambda = -\omega^2$ can be computed explicitly as,
\begin{eqnarray*}
\lambda_{1} &=& \frac{\omega_a + \omega_d + \sqrt{ \rule{0pt}{.4cm} (\omega_a + \omega_d)^2 - 4(\omega_a\omega_d -\omega_b \omega_c )  }   }{2} \\
\lambda_{2} &=& \frac{\omega_a + \omega_d - \sqrt{ \rule{0pt}{.4cm} (\omega_a + \omega_d)^2 - 4(\omega_a\omega_d -\omega_b \omega_c )  }   }{2}
\end{eqnarray*}
with associated eigenvectors $v_1$ and $v_2$. This results in four frequencies
\begin{equation}
\omega_{ \pm 1}(k,\ell) =  \pm \sqrt{- \lambda_1}, \qquad \omega_{\pm 2}(k,\ell) =  \pm \sqrt{- \lambda_2}. \label{eigs}
\end{equation}
 See Fig.~\ref{fig:spectrum} for an example plot of the dispersion surfaces.
\begin{figure}
    \centering
  \begin{tabular}{@{}p{0.33\linewidth}@{\quad}p{0.33\linewidth}@{\quad}p{0.33\linewidth}@{}}
    \subfigimg[width=\linewidth]{\bf (a)}{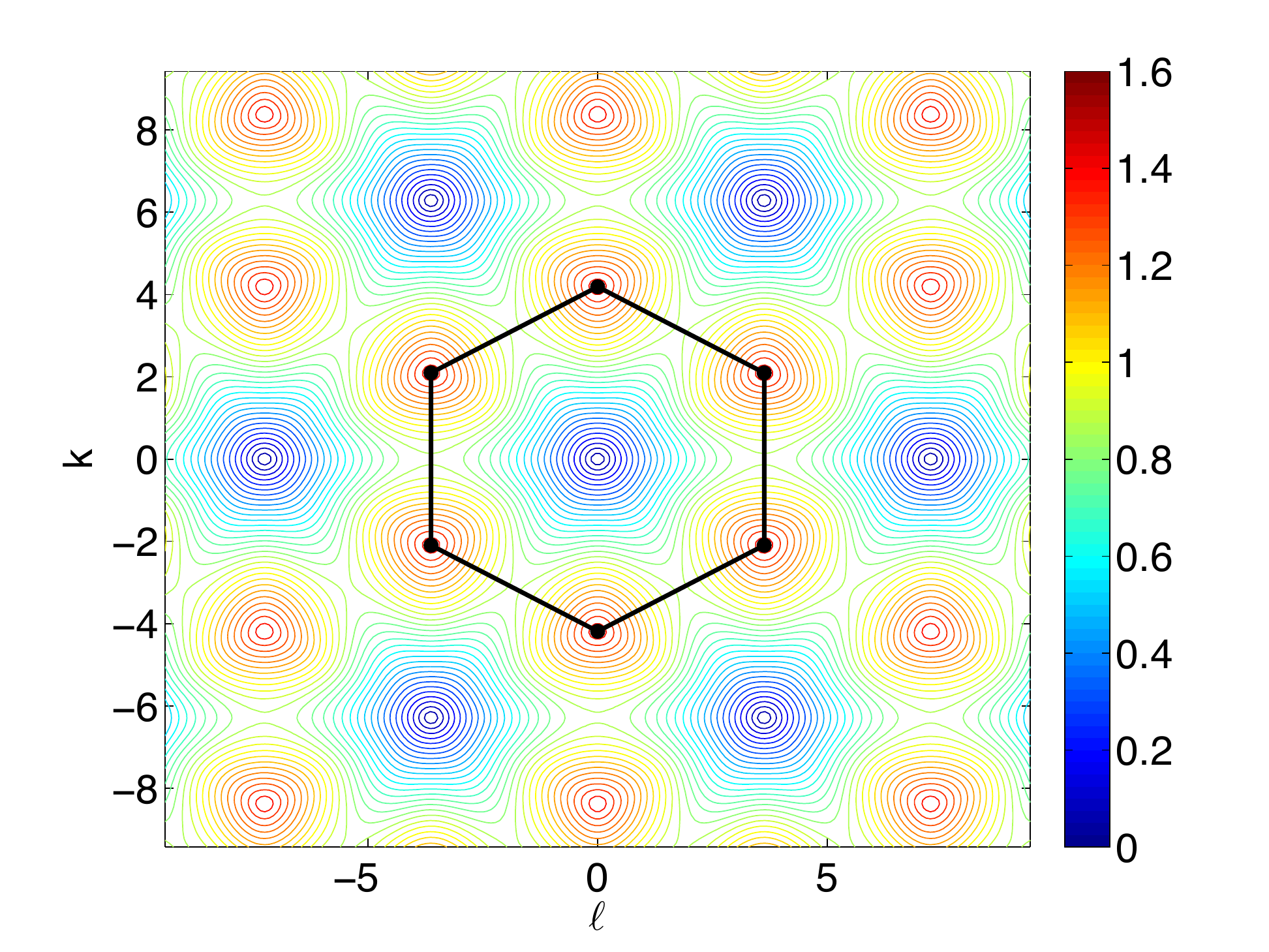} &
     \subfigimg[width=\linewidth]{\bf (b)}{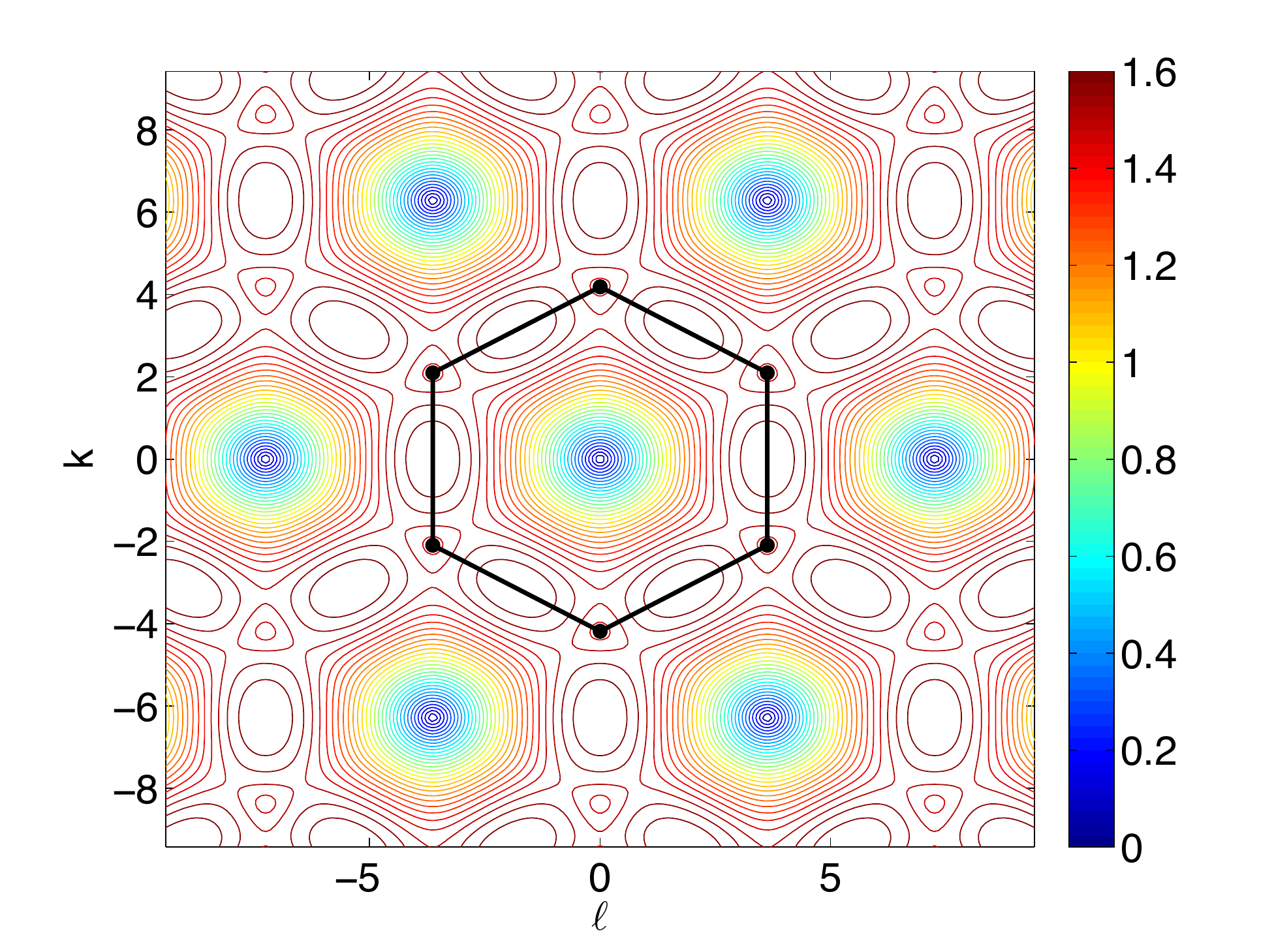} &
    \subfigimg[width=\linewidth]{\bf (c)}{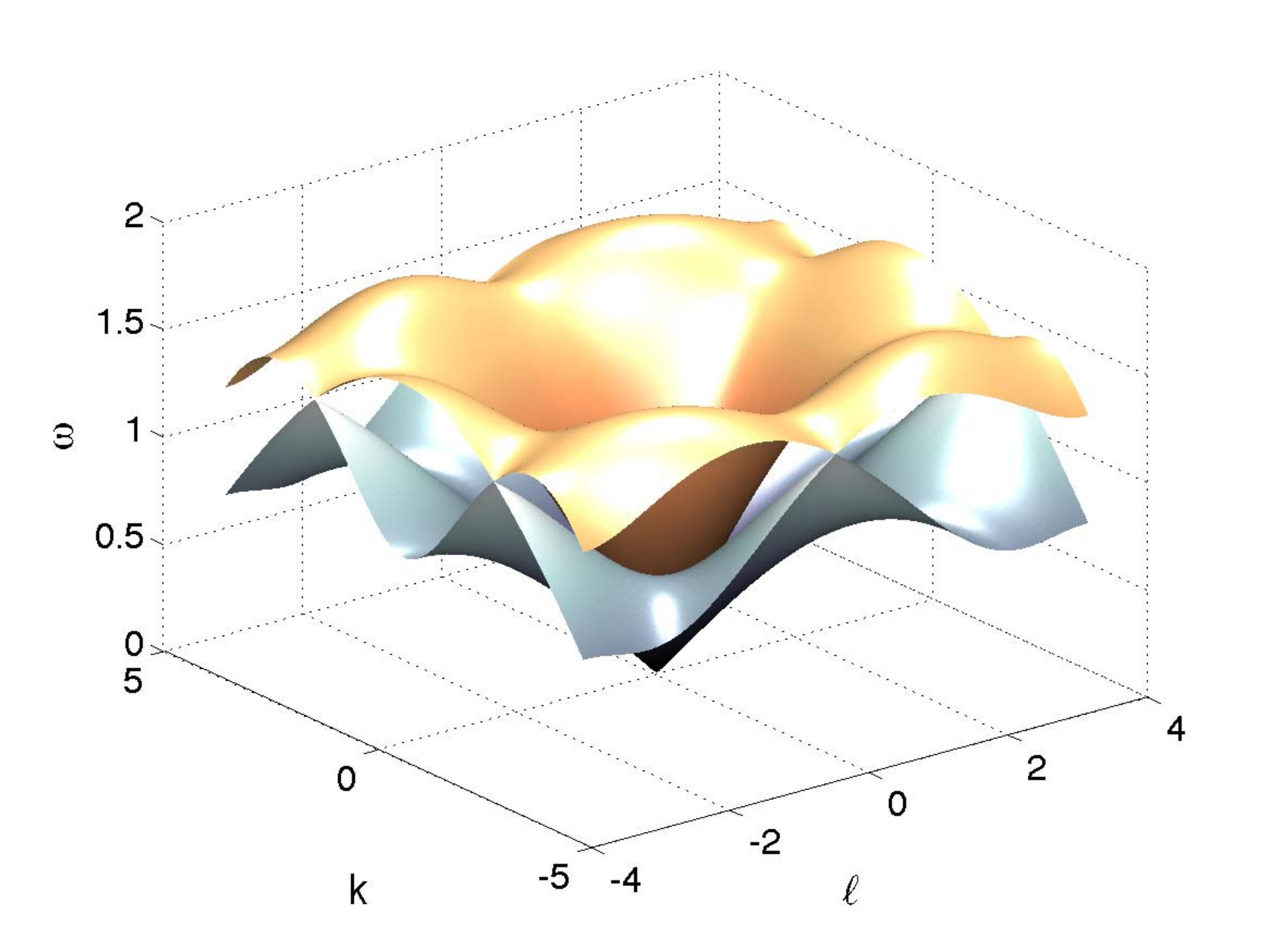}
  \end{tabular}
 \caption{The dispersion surface $\omega(k,\ell)$ as given by Eq.~\eqref{eigs}. The parameter values are $d=M=\gamma=1$
 and $\delta = 0.1$.  \textbf{(a)} Contour plot of bottom dispersion surface. The Brillouin zone is shown as a black line and the Dirac points are given
 as black points. \textbf{(b)} Same as the left, but for the top dispersion surface. \textbf{(c)} Both dispersion surfaces in ($k,\ell,\omega$) space.}
 \label{fig:spectrum}
\end{figure}
From inspection of Fig.~\ref{fig:spectrum}, one can see regions where the top and bottom dispersion surfaces form a downward and upward pointing cone respectively; these are the Dirac cones. The point where these
two cones meet is the Dirac point. To explicitly calculate the location of the Dirac points, we find values of $(k,\ell)$ where the two surfaces meet. In the case of Eq.~\eqref{eigs}, the relevant condition is,
\begin{equation} \label{eq:cond0}
0 =  (\omega_a + \omega_d)^2 - 4(\omega_a\omega_d -\omega_b \omega_c ).
\end{equation}
Direct inspection of the linear coefficients reveals that
\begin{align*}
 a_2 = a_3, \qquad d_2=d_3, \qquad b_1=c_1=0,  \qquad  b_2 = c_2  = -b_3 = -c_3,
\end{align*}
and,
\begin{equation} \label{eq:rel}
\frac{a_1 - d_1}{a_2 - d_2} = -2.
\end{equation}
Thus, we have that
\begin{eqnarray*}
\omega_a(k,\ell) &=&   4 a_1 \sin^2\left( k/2 \right)     +    4 a_2 \left( 1 - \cos(k/2)\cos(\sqrt{3}/2 \, \ell)       \right) \\
\omega_d(k,\ell) &=&   4 d_1 \sin^2\left( k/2 \right)     +    4 d_2 \left( 1 - \cos(k/2)\cos(\sqrt{3}/2 \, \ell)       \right) \\
\omega_c(k,\ell)&=&\omega_b(k,\ell) = 4 b_2 \sin(k/2)\sin(\sqrt{3}/2  \ell)
\end{eqnarray*}
With these simplifications Eq.~\eqref{eq:cond0} becomes,
\begin{equation}
0=(\omega_a - \omega_d)^2 + 4 \omega_b^2
\end{equation}
which is equivalent to the set of equations,
\begin{subequations} \label{eq:cond}
\begin{eqnarray}
- \frac{a_1 - d_1}{a_2 - d_2} &=& \frac{ 1 - \cos(k/2)\cos\left(\sqrt{3}/2 \, \ell \right)}{ \sin^2\left( k/2 \right)} \\ [2ex]
 0 &=&   \sin(k/2)\sin(\sqrt{3}/2  \ell).
\end{eqnarray}
\end{subequations}
Equation~\eqref{eq:cond} along with Eq.~\eqref{eq:rel} reveals six non-trivial solution points, which, as expected, are situated on the vertices of the Brillouin zone:
$(k_d,\ell_d) =  (\pm \, 4\pi/3, 0) $ and $(k_d, \ell_d) =  (\pm \, 2\pi/3, \pm \, 2 \pi / \sqrt{3} ) $.
We now verify that the shape of the dispersion surface is conical near the Dirac point. To that end, we Taylor expand the functions $\omega_a,\omega_b,\omega_c$ and $\omega_d$ about
the Dirac point $(k_d,\ell_d)$, substitute into Eq.~\eqref{eigs} and ignore higher order terms, yielding
\begin{equation}  \label{eq:omg_order1}
\omega(k,\ell) \approx \pm \sqrt{   \omega_{cr} ^2 \pm   \sqrt{  3 (a_1-a_2)^2 (k-k_d)^2 + 9 b_2^2(l-l_d)^2    }    }
\end{equation}
where $ \omega_{cr} = \omega(k_d,l_d)$ is the frequency at the Dirac point. Making once again use of the symmetries of the system,
one can show via a direct calculation that $a_1-a_2 = \sqrt{3} b_2$.
Defining the polar coordinates $\eta \cos(\theta) = k - k_d$,
$\eta \sin(\theta) = \ell - \ell_d$ and Taylor expanding about $\eta = 0$ yields
\begin{equation}
\omega(k,\ell)  \approx \pm ( \omega_{cr} \pm \alpha \eta ) = \pm \left( \omega_{cr} \pm \alpha \sqrt{ (k-k_d)^2 + (l-l_d)^2     }  \right)
\end{equation}
Thus, to first order, the dispersion surfaces form upward and downward pointing cones  with slope $\alpha = \frac{3 b_2}{2 \omega_{cr}} $  that meet at the Dirac point $\omega_{cr}$; a zoom into a particular
case example is shown in Fig.~\ref{fig:slice}(a).

Equation~\eqref{eq:cond} along with Eq.~\eqref{eq:rel} also admits a ``trivial'' solution $(k_0,\ell_0)=(0,0)$ which is located at the center of the Brillouin zone. Near this point the functions $\omega_{a,b,c,d}$ have the leading order expansions
\begin{equation}
\omega_a=a_1k^2+a_2(k^2+3\ell^2)/2,\quad \omega_d=d_1k^2+d_2(k^2+3\ell^2)/2,\quad \omega_b=\omega_c=\sqrt{3}b_2k\ell,
\end{equation}
or in terms of $\hatd$ and $\hatdelta$
\begin{equation}
\omega_a+\omega_d=\frac{1}{2}(5\hatdelta-3\hatd)\frac{3(k^2+\ell^2)}{2},\quad
\omega_a-\omega_d=\frac{1}{4}(-\hatdelta+3\hatd)\frac{3(-k^2+\ell^2)}{2},\quad
\omega_b=\omega_c=\frac{3}{8}(\hatdelta-3\hatd)k\ell.
\end{equation}
Therefore the dispersion surface is again locally conical, but this time consists of four cones with two group speeds $\alpha_\pm$:
\begin{equation}
\omega^2=\alpha_\pm^2(k^2+\ell^2),\quad \alpha_\pm^2=\frac{3}{4}\left(\frac{1}{2}(3\hatd-5\hatdelta)\pm\frac{1}{4}(3\hatd-\hatdelta)\right);
\end{equation}
see Fig.~\ref{fig:slice} (b) for a zoomed-in view.

In the linear limit $0<\hatdelta/\hatd=\delta/d\ll1$, these two group speeds are respectively $\alpha_-=3\sqrt{\hatd}/4$ and $\alpha_+=\sqrt{3}\alpha_-$. In the nonlinear regime, while $\alpha_\pm$ are both purely real for $\delta/d<1/3$, $\alpha_-$ and $\alpha_+$ respectively become purely imaginary for $\delta/d>1/3$ and $\delta/d>9/11$. The existence of purely imaginary eigenvalues implies that the system is linearly unstable in this long wave regime with saturation
arising in the nonlinear regime. Examples of stable and unstable propagation are considered via numerical simulations in Sec.~\ref{sec:numerics}.
It is worth noting that experimentally relevant values of the precompression would typically satisfy these stability conditions.

\begin{figure}[h]
    \centering
  \begin{tabular}{@{}p{0.4\linewidth}@{\quad}p{0.4\linewidth}@{}}
    \subfigimg[width=\linewidth]{\bf (a)}{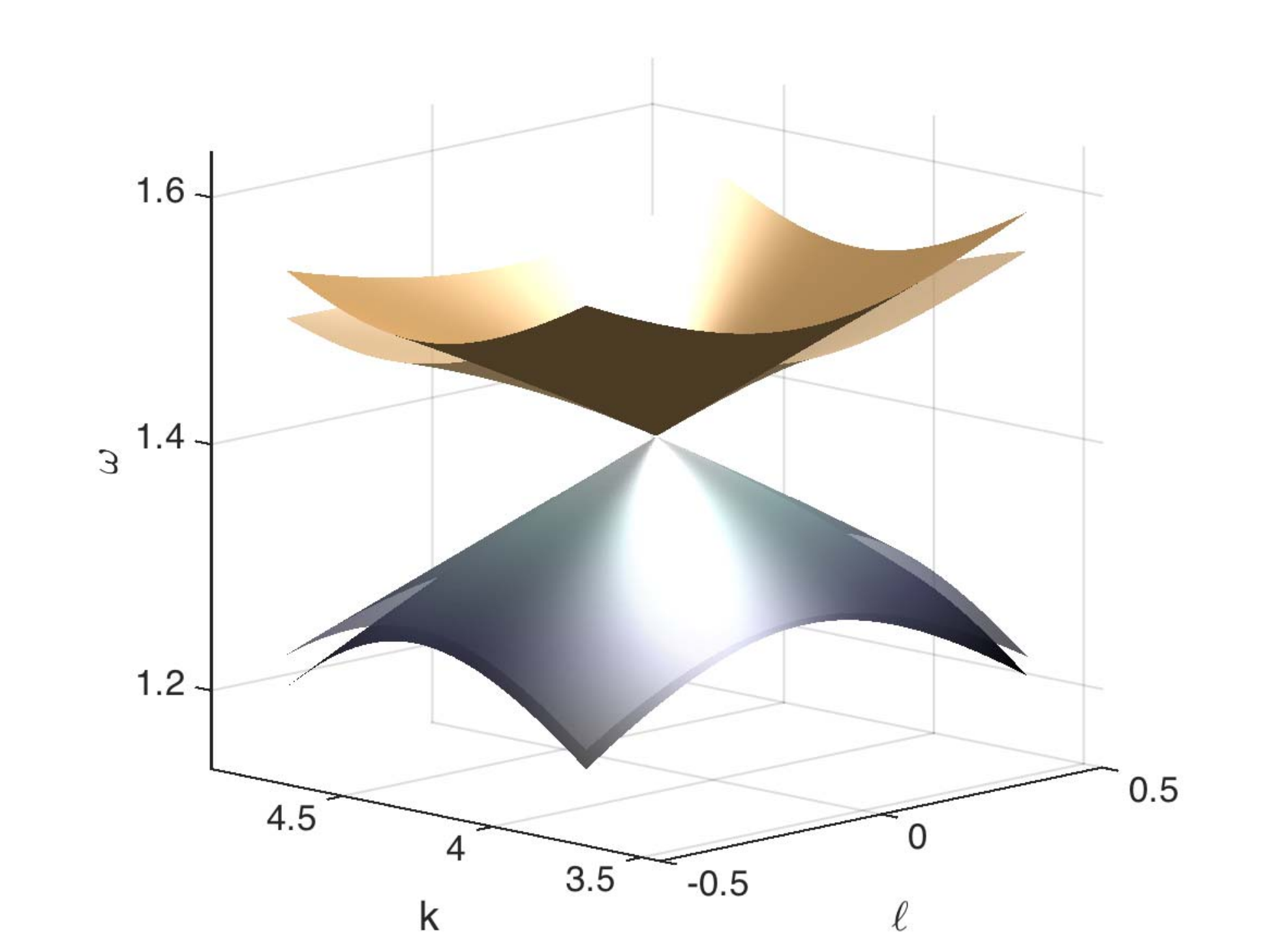} &
     \subfigimg[width=\linewidth]{\bf (b)}{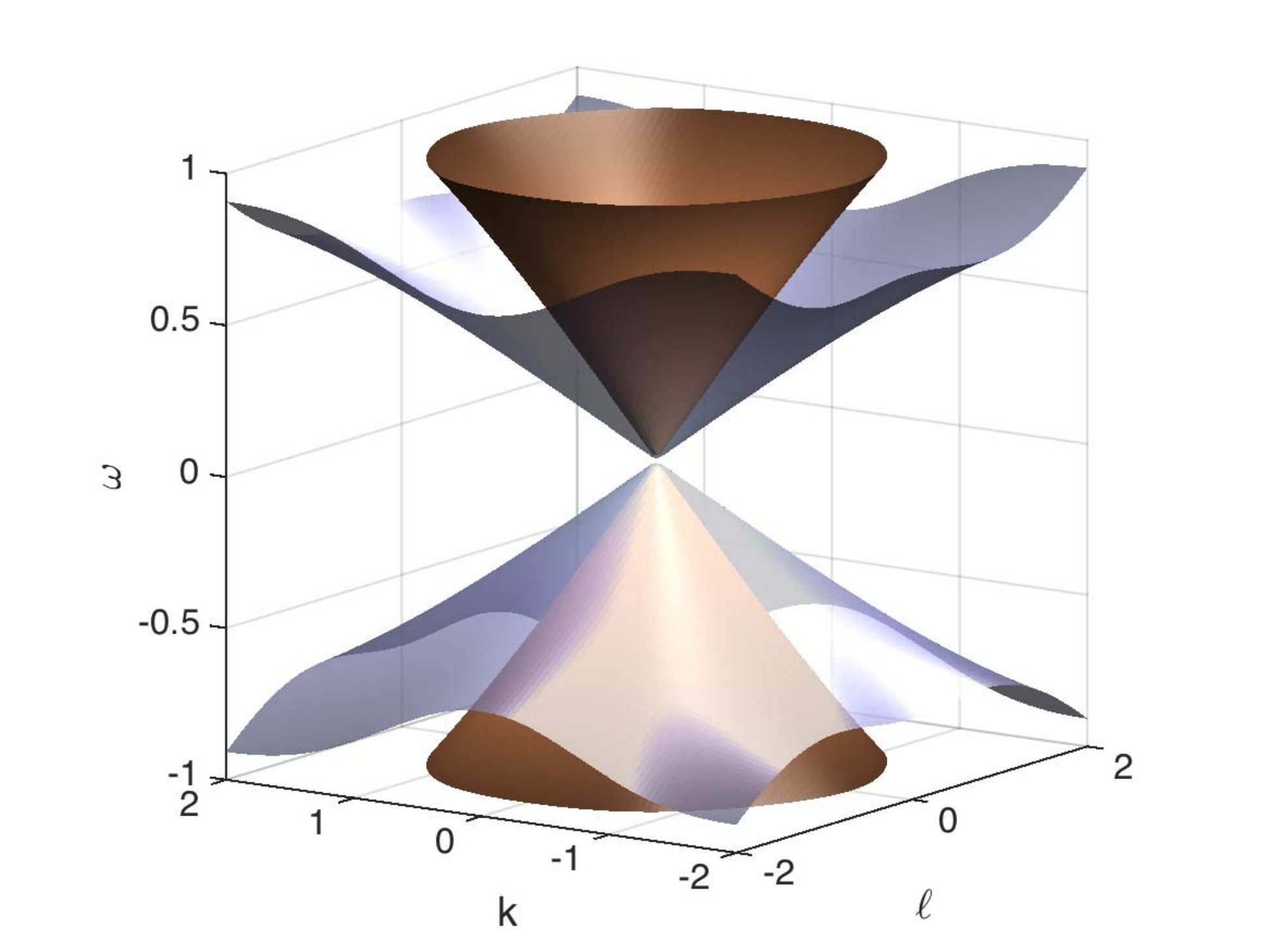}
  \end{tabular}
  \caption{\textbf{(a)} Zoom-in of the dispersion surface near the Dirac point $(k_d,\ell_d) = (4 \pi/3,0)$. The transparent layers are the first order approximations given by $\omega_{cr} \pm \alpha \sqrt{ (k-k_d)^2 + (l-l_d)^2 }$
  where the slope is  $\alpha = 3 b_2 / 2 \omega_{cr}. $ \textbf{(b)} Zoom-in of the dispersion surface near the Dirac point $(k_0,\ell_0) = (0,0)$. Here, there are four cones meeting
  at the Dirac point rather than two, as in panel (a). Note that
no first order approximation is shown in panel (b).   }
  \label{fig:slice}
 \end{figure}

\subsection{Heuristic argument for conical diffraction}

It has been shown in various contexts that conical diffraction is possible in systems with Dirac cones in the dispersion relation \cite{Hamilton,Peleg,Ablowitz1}. A suitable  initial
condition in order to observe the relevant phenomenology
is a localized function (such as a Gaussian) that is modulating a Bloch
wave with a wavenumber near the Dirac point. The resulting
conical diffraction evolution
will dynamically yield an expanding ring
with constant width and amplitude around the ring (however, the amplitude
decays as the ring expands).  In this subsection we present a
heuristic calculation that partially explains why such an initial condition can lead to conical diffraction. The calculation is adapted from \cite{Peleg} for a discrete system.
The general solution of the linearized equation Eq.~\eqref{eq:linear} has the form
 $$   x_{m,n}(t) =   \sum_{j \in \{-2,-1,1,2 \} } I_j $$
 for the $x_{m,n}$ component (and likewise for the $y_{m,n}$ component) where we make use of the inverse transform
of Eq.~\eqref{eq:DFT},

$$ I_j  = \frac{1}{2\sqrt{3} (2\pi)^2} \int_{- 2 \pi / \sqrt{3}}^{2 \pi / \sqrt{3}} \int_{0}^{2 \pi}  \exp\left( - i(k m + \frac{n}{2} ( k + \l \sqrt{3})  )  \right)
     C_j(k,\l, v_{|j|}   )  e^{ i \omega_j(k,\l) t }  \, dk \, dl   $$
where $\omega_j(k,\l)$ is given by Eq.~\eqref{eigs}. The coefficients $C_j$ depend on the initial data $\mathbf{q}_{m,n}(0)$, wavenumbers $(k,\ell)$, and eigenvectors $v_1,v_2$.
We pick an initial displacement that is a localized function (such as a Gaussian) modulating a Bloch mode with a wavenumber pair $(k_d,\ell_d)$ that is near the Dirac point.
Since the dispersion surface is cone-like near the Dirac point, we have
$\omega(k,\ell) \approx \alpha \sqrt{ k^2 + \ell^2}$, where $\alpha = 3b_2/2 \omega_{cr}$, as we saw above. For notational simplicity, we have made a change of variable to shift the cone to the origin.
With these assumptions, we can write

 $$   I_j  \approx \frac{1}{2\sqrt{3} (2\pi)^2} \int_{- 2 \pi / \sqrt{3}}^{2 \pi / \sqrt{3}} \int_{0}^{2 \pi}  \exp\left( - i(k m + \frac{n}{2} ( k + \l \sqrt{3}) )  \right)
   C_j(k,\l, v_{|j|}   ) \exp \left( \alpha \, i \, \sqrt{ k^2 + \ell^2} \, t  \right)  \, dk \, dl   $$
If we assume that the initial amplitude is radially symmetric in Fourier space
and  is compactly supported in a circle of radius $\pi$,
then we can make use of the polar coordinates $k = \eta \cos(\theta)$ , $\ell = \eta \sin(\theta)$
in order to rewrite the above integral as,
 $$    I_j \approx   \frac{1}{2\sqrt{3} (2\pi)^2}  \int_{0}^{\pi}   C_j( \eta, v_{|j|}   ) \exp( i \alpha \eta t )  \eta  \int_{-\pi}^{\pi}   \exp\left( - i \eta(   \cos(\theta) m + \frac{n}{2} ( \cos(\theta) + \sin(\theta) \sqrt{3} ) ) \right)        \, d\theta \, d\eta.   $$
If we use the identity
 \begin{equation}
 \cos(\theta) (m + n/2)  + \sin(\theta) n \sqrt{3}/2  = \sqrt{m^2 + n^2  + n m} \, \sin(\theta + \phi),
 \end{equation}
 where $ \displaystyle \phi = \tan^{-1}\left( \frac{m+n/2}{ n \sqrt{3}/2 } \right)$ then we have
 $$    I_j \approx \frac{1}{2\sqrt{3} (2\pi)^2}  \int_{0}^{\pi} C_j( \eta, v_{|j|}   ) \exp( i \alpha \eta t ) \eta   \int_{-\pi}^{\pi}  e^{- i \eta \rho \sin(\theta)}    d\theta d\eta   $$
 where we dropped the phase shift $\phi$ since the second integration is over an entire period and where we defined $\rho = \sqrt{n^2 + m^2 + nm}$.  Note
 that within the hexagonal coordinate frame, the expression $\rho$ is radially symmetric.
 The second integral is a zeroth order Bessel function
 and so we have
 $$  I_j \approx \frac{1}{4 \sqrt{3} \pi}   \int_{0}^{\pi} C_j( \eta, v_{|j|}   ) \exp( i \alpha \eta t )    \eta  J_0(\eta \rho)   d\eta   $$
 For Gaussian initial data there is no closed form
expression for this integral. However, if we assume that each component of the initial data $\textbf{q}(\eta,0)$ has the form of an
 exponential with decay rate $g>0$ and that the eigenvectors do not vary much in the vicinity of the Dirac point then $C_j$ will be a linear combination of exponential functions, which we write
 as  $C_j = B_j( v_1(k_d,\l_d), v_2(k_d,\l_d) ) e^{- g \eta}$. Therefore, we have
 $$   I_j \approx \frac{B_j}{4 \sqrt{3} \pi}  \int_{0}^{\infty}   \exp( -\eta(g - i \alpha t) )   \eta  J_0(\eta \rho)   d\eta     - \frac{B_j}{4 \sqrt{3} \pi}   \int_{\pi}^{\infty} \exp( -\eta(g - i \alpha t) )    \eta  J_0(\eta \rho)   d\eta    $$
 Assuming the contribution of the second integral is small with respect to the first (which can be computed formally
 by setting $s = g - i \alpha t$ and computing the Laplace transform of the function $ f(\eta) = J_0(\eta \rho) \eta$), we finally have a closed form approximation of the integral
 \begin{equation} \label{eq:ideal}
     I_j \approx  \frac{B_j}{4 \sqrt{3} \pi}  \frac{ (g - i \alpha t)  }{    ( (g - i \alpha t)^2 + \rho^2 )^{3/2}    }
 \end{equation}
 The solution $x_{m,n}(t)$ will be a linear combination of the real and imaginary parts of the $I_j$, where each has the form of an expanding ring
 as $t$ increases for all $\alpha t \gg \rho$.
 Note the key aspect to this calculation
 was representing the dispersion surface $\omega(k,\ell)$ as $\alpha \sqrt{k^2 + l^2}$, which, along with the fact the initial condition is localized in Fourier space,
 allowed us to write an approximate linear solution in terms of a radially symmetric function in physical space.

While there were numerous heuristic assumptions in our calculation
above, we also now give a more systematic asymptotic analysis of
the linear dynamics, utilizing a multiple
scales expansion in the vicinity of the Dirac point, in order
to complement and corroborate the above argument.

\section{Derivation of a Dirac System}\label{sec:dirac_eq}

We start by considering the multiple-scale ansatz:
\begin{equation} \label{eq:ansatz}
\q_{m,n}(t) =  \epsilon \phi_P(T) E  + c.c., \qquad E = e^{ i( k m + \frac{n}{2}(k + \sqrt{3} \ell))} e^{i \omega t},  \quad  P = \epsilon\left( m + \frac{n}{2}, \frac{n \sqrt{3}}{2}   \right), \quad  T = \epsilon t
\end{equation}
where $ \phi_P(T) \in \C^2$ .
Here, $E$ represents a plane wave, $\phi$ the slow envelope, while
$P$ and $T$ represent the slow evolution variables in space and time
respectively.
Substitution of the above ansatz into Eq.~\eqref{eq:linear} and ignoring $\mathcal{O}(\epsilon^3)$ terms and higher yields,
\begin{eqnarray*}
&& \epsilon^2 2 i \omega \frac{d \phi_{P}}{dT}  - \epsilon \phi_P \omega^2   = \epsilon 2(DF_1 + DF_2 +DF_3 ) \phi_{P} \\
&&  -\epsilon(  DF_1(  \phi_{ P+ \epsilon v_1} e^{i \kappa \cdot v_1}  + \phi_{P - \epsilon v_1} e^{-i \kappa \cdot v_1} )
+ DF_2(  \phi_{P+ \epsilon v_2} e^{i \kappa \cdot v_2}  + \phi_{P- \epsilon v_2} e^{-i \kappa \cdot v_2} )
+DF_3(  \phi_{P+ \epsilon v_3} e^{i \kappa \cdot v_3}  + \phi_{P - \epsilon v_3} e^{-i \kappa \cdot v_3} )
)
\end{eqnarray*}
where $\kappa = (k,\ell)$ and $v_1 = (1,0)$, $v_2 = (1/2,\sqrt{3}/2)$, $v_3 = (1/2,-\sqrt{3}/2)$. We now make use of the Taylor series expansion,
$$  \phi_{P \pm \epsilon v_j} \approx \phi_{P} \pm \epsilon D\phi_P v_j, \qquad j=1,2,3   $$
where
$$ \phi_P = \phi(X,Y) = \begin{pmatrix}  \alpha(X,Y) \\ \beta(X,Y) \end{pmatrix}, \qquad  D\phi = \begin{pmatrix}  \partial_X \alpha &  \partial_Y \alpha \\ \partial_X \beta &  \partial_Y \beta \end{pmatrix} $$
which yields,
\begin{eqnarray*}
&&\epsilon^2 2 i \omega \frac{d \phi_{P}}{dT}  - \epsilon \phi_P \omega^2   = \epsilon 2(DF_1 + DF_2 +DF_3 ) \phi_{P}
 -\epsilon(  DF_1(     (\phi_P + \epsilon D\phi_P v_1)  e^{i \kappa \cdot v_1}  + (\phi_P - \epsilon D\phi_P v_1) e^{-i \kappa \cdot v_1} )  \\
 && + DF_2( (\phi_P + \epsilon D\phi_P v_2)e^{i \kappa \cdot v_2}  + (\phi_P - \epsilon D\phi_P v_2) e^{-i \kappa \cdot v_2} )
+DF_3(  (\phi_P + \epsilon D\phi_P v_3) e^{i \kappa \cdot v_3}  + (\phi_P - \epsilon D\phi_P v_3)e^{-i \kappa \cdot v_3} ) )
)
\end{eqnarray*}
 Then, at order $\epsilon$ we have
\begin{equation} \label{eb}
 0 = - \omega^2 \phi_P -    2( DF_1 + DF_2 +DF_3  -   DF_1 \cos( \kappa \cdot v_1 ) -  DF_2 \cos(\kappa \cdot v_2)  - DF_3 \cos( \kappa \cdot v_3) )\phi_P,
\end{equation}
while at order $\epsilon^2$ we have
\begin{equation}  \label{dirac}
  i \omega \frac{d \phi_{P}}{dT}  =   DF_1   \left( D \phi_P \, v_1\right)  i \sin(  \kappa \cdot v_1 ) + DF_2 (D \phi_P \, v_2 )i \sin(\kappa \cdot v_2)  + DF_3 (D\phi_P \, v_3) i \sin(\kappa \cdot v_3).
\end{equation}
The solutions to these linear PDEs have the form $\phi_P(T) = \phi(X,Y,T) = \tilde{v} e^{ i(\tilde{k} X + \tilde{\ell} Y + \mu T)}$. In Fourier space, Eq.~(\ref{eb}) and (\ref{dirac}) respectively become Eq.~(\ref{eq:disp_matrix}) at $\mathcal{O}(1)$ and $\mathcal{O}(\epsilon)$. The former requires that $\omega^2=\omega_{cr}^2$. If we evaluate the wavenumber at the Dirac point, the latter becomes a linear Dirac equation. For example, if we take $\kappa = ( 4 \pi /3, 0)$, then we have
\begin{equation}
-2 \mu \omega \tilde{v} =  \widetilde{\mathcal{H}} \tilde{v}, \qquad   \widetilde{\mathcal{H}} :=  \begin{pmatrix}    \tilde{k} \sqrt{3} ( a_2 - a_1) &  3 b_2 \tilde{\ell} \\  3 b_2 \tilde{\ell}  & \tilde{k}  \sqrt{3}(d_2 - d_1) \end{pmatrix},
\end{equation}
or back to physical space
\begin{equation}
\partial_T\phi_P(T) = \pm\frac{1}{2\omega_{cr}} \begin{pmatrix}    \sqrt{3} ( a_2 - a_1) \partial_X &  3 b_2 \partial_Y \\  3 b_2 \partial_Y  & \sqrt{3}(d_2 - d_1) \partial_X \end{pmatrix} \phi_P(T).
\end{equation}
The dispersion relation of the Dirac equation is
$$ \mu =  \pm \frac{ 3 b_2 }{2\omega_{cr}} \sqrt{ \tilde{k}^2 + \tilde{l}^2 } $$
}
where we made use of the fact that $-(d_1-d_2) = a_1-a_2 = \sqrt{3} b_2$. Note the connection to the approximation given in ~Eq.~\eqref{eq:omg_order1}
through the relation $\omega \approx \pm ( \omega_{cr}  + \mu)$.

Near $(k_0,\ell_0)=(0,0)$, the envelope equation can be obtained by expanding Eq.~(\ref{eq:linear}) to $\mathcal{O}(\epsilon^3)$. Alternatively, we can expand the dispersion relation (\ref{eq:disp_matrix}) to $\mathcal{O}(\epsilon^2)$ and replace $(k,\ell,\omega)$ by $-i\epsilon(\partial_X,\partial_Y,\partial_T)$. The resulting envelope equation is the following vector wave equation:
\begin{equation}
\partial_{TT}\phi_P(T)=-\left(
\begin{array}{cc}
a_1\partial_{XX}+a_2(\partial_{XX}+3\partial_{YY})/2 & \sqrt{3}b_2\partial_{XY}\\
\sqrt{3}b_2\partial_{XY} & d_1\partial_{XX}+d_2(\partial_{XX}+3\partial_{YY})/2
\end{array}
\right)\phi_P(T).
\end{equation}

We now turn to numerical computations in order to explore the validity
of the above considerations in the linear limit, as well as to extend
them in the nonlinear regime.

\section{Numerical study of the transition between linear  and nonlinear conical wave propagation} \label{sec:numerics}

\begin{figure}
     \centering
   \begin{tabular}{@{}p{0.33\linewidth}@{\quad}p{0.33\linewidth}@{\quad}p{0.33\linewidth}@{}}
    \subfigimg[width=\linewidth]{\bf (a)}{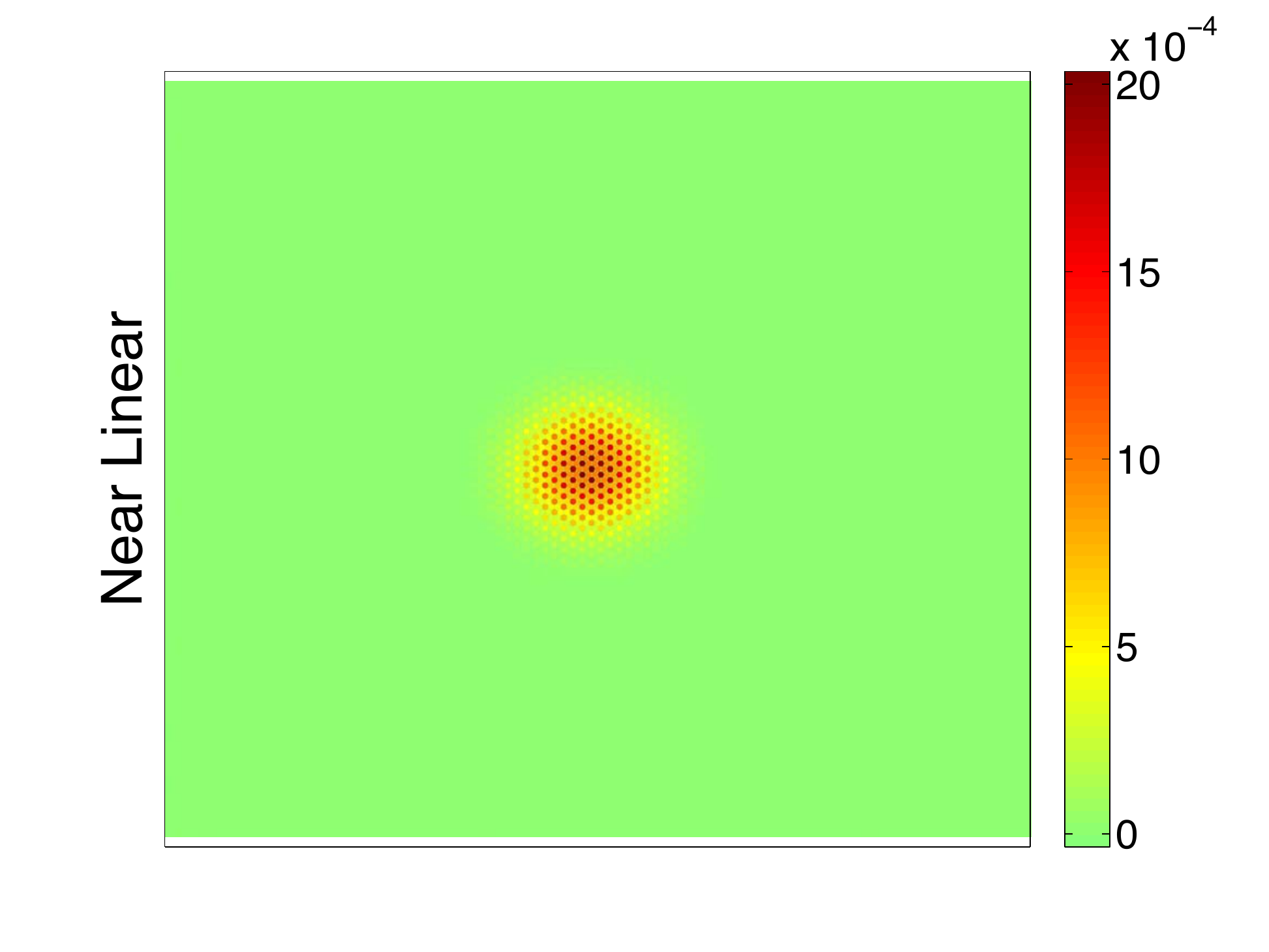} &
     \subfigimg[width=\linewidth]{\bf (b)}{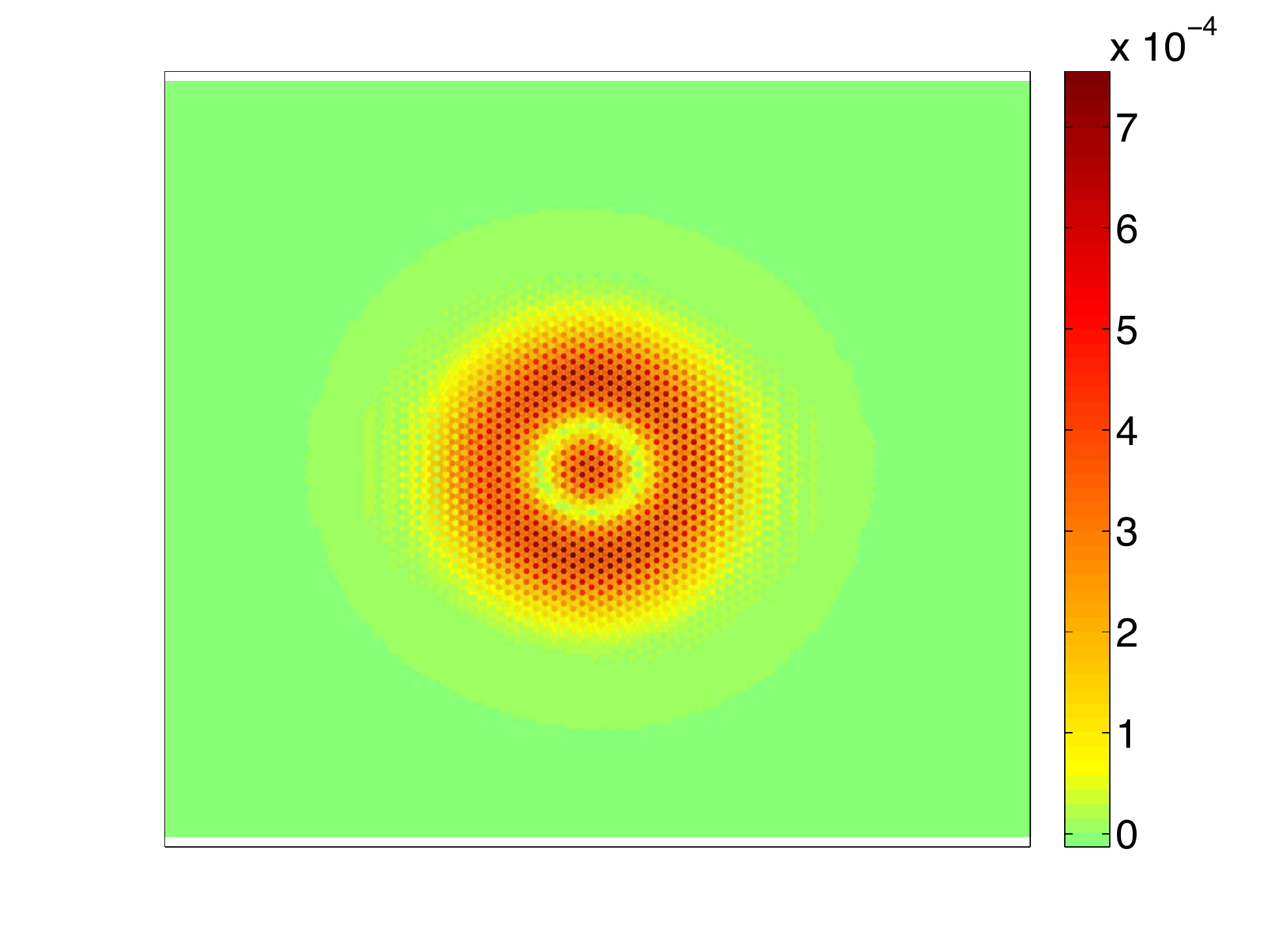} &
    \subfigimg[width=\linewidth]{\bf (c)}{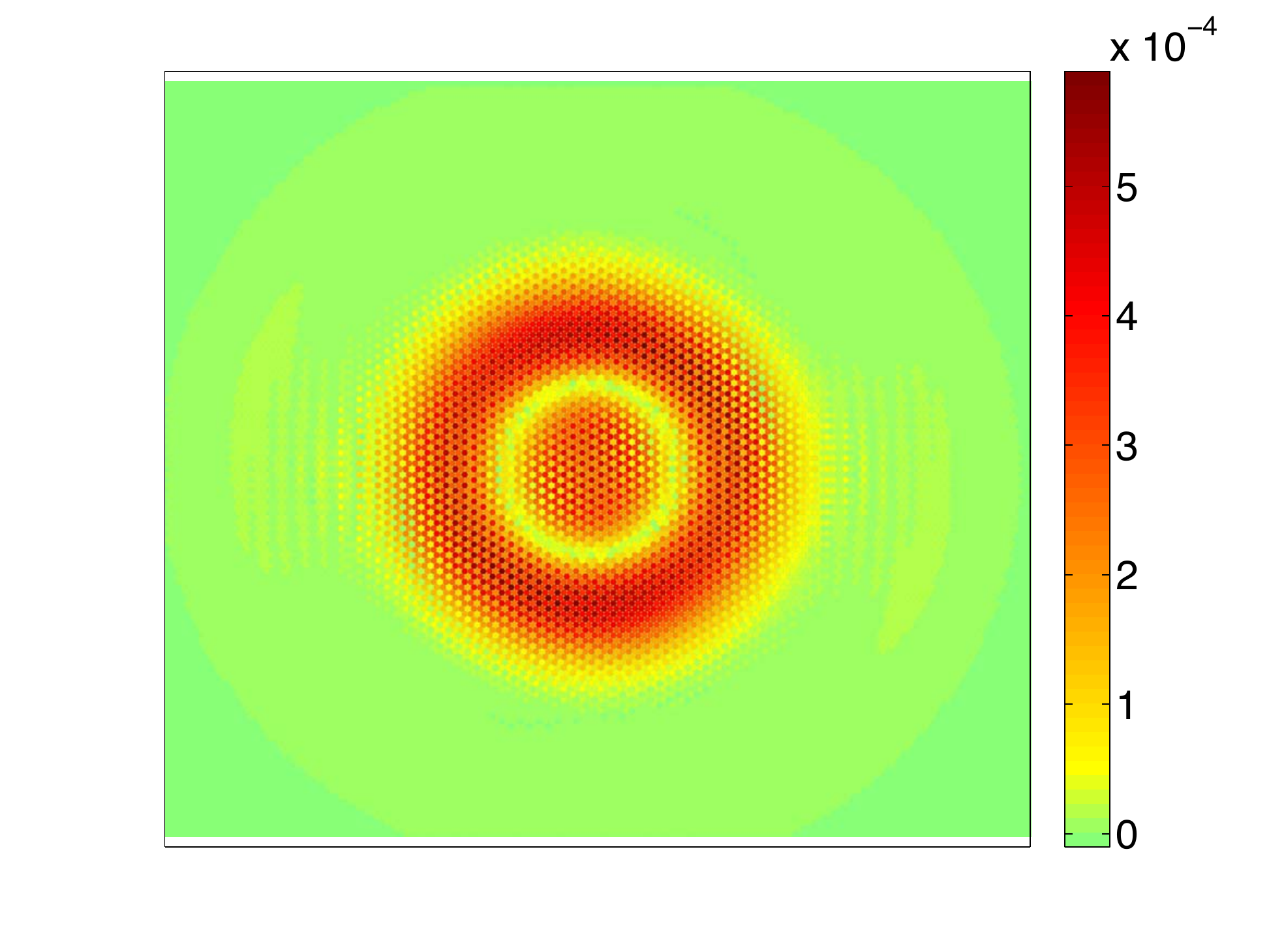}
  \end{tabular}
   \begin{tabular}{@{}p{0.33\linewidth}@{\quad}p{0.33\linewidth}@{\quad}p{0.33\linewidth}@{}}
    \subfigimg[width=\linewidth]{\bf (d)}{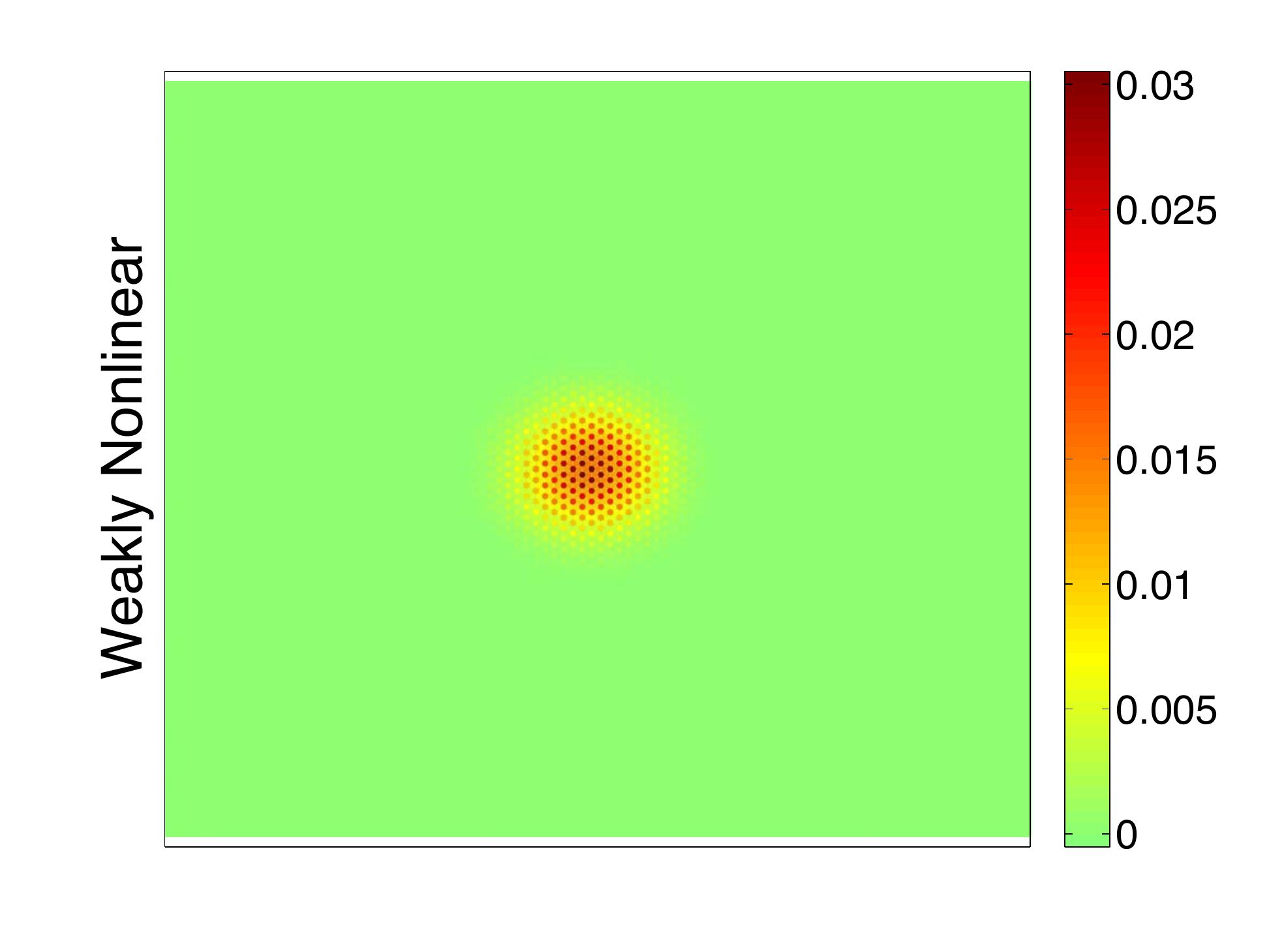} &
     \subfigimg[width=\linewidth]{\bf (e)}{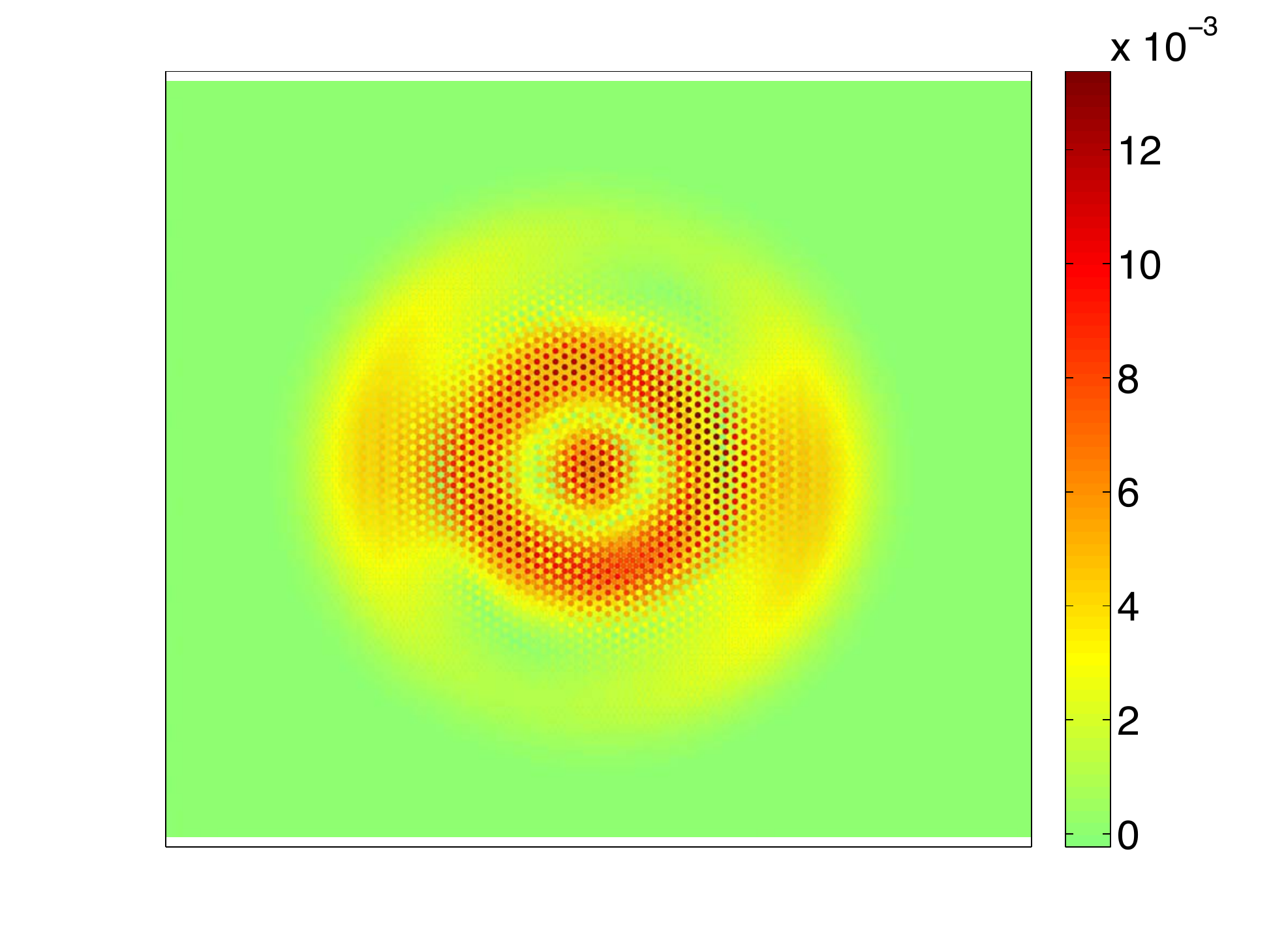} &
    \subfigimg[width=\linewidth]{\bf (f)}{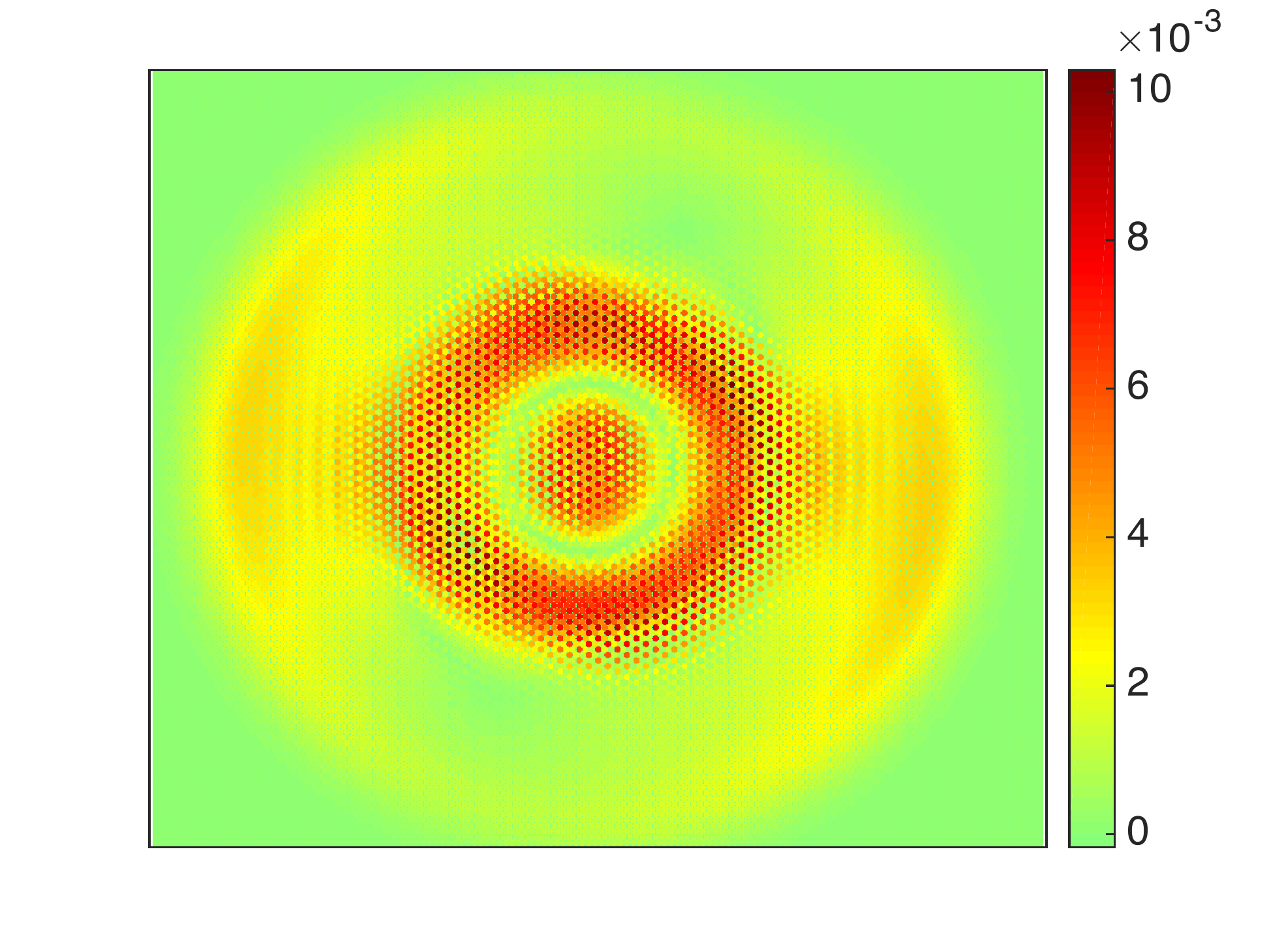}
  \end{tabular}
     \begin{tabular}{@{}p{0.33\linewidth}@{\quad}p{0.33\linewidth}@{\quad}p{0.33\linewidth}@{}}
    \subfigimg[width=\linewidth]{\bf (g)}{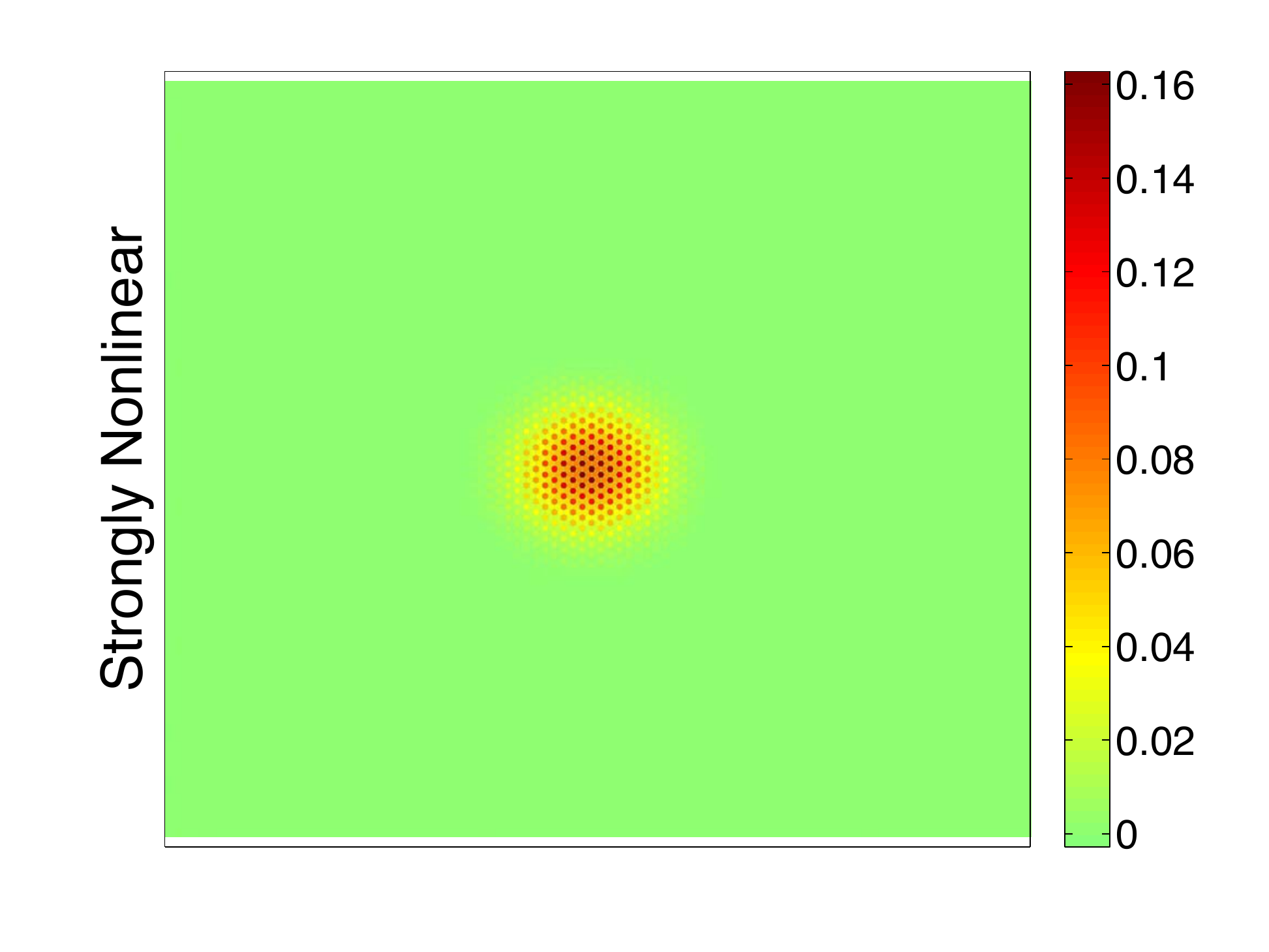} &
    \subfigimg[width=\linewidth]{\bf (h)}{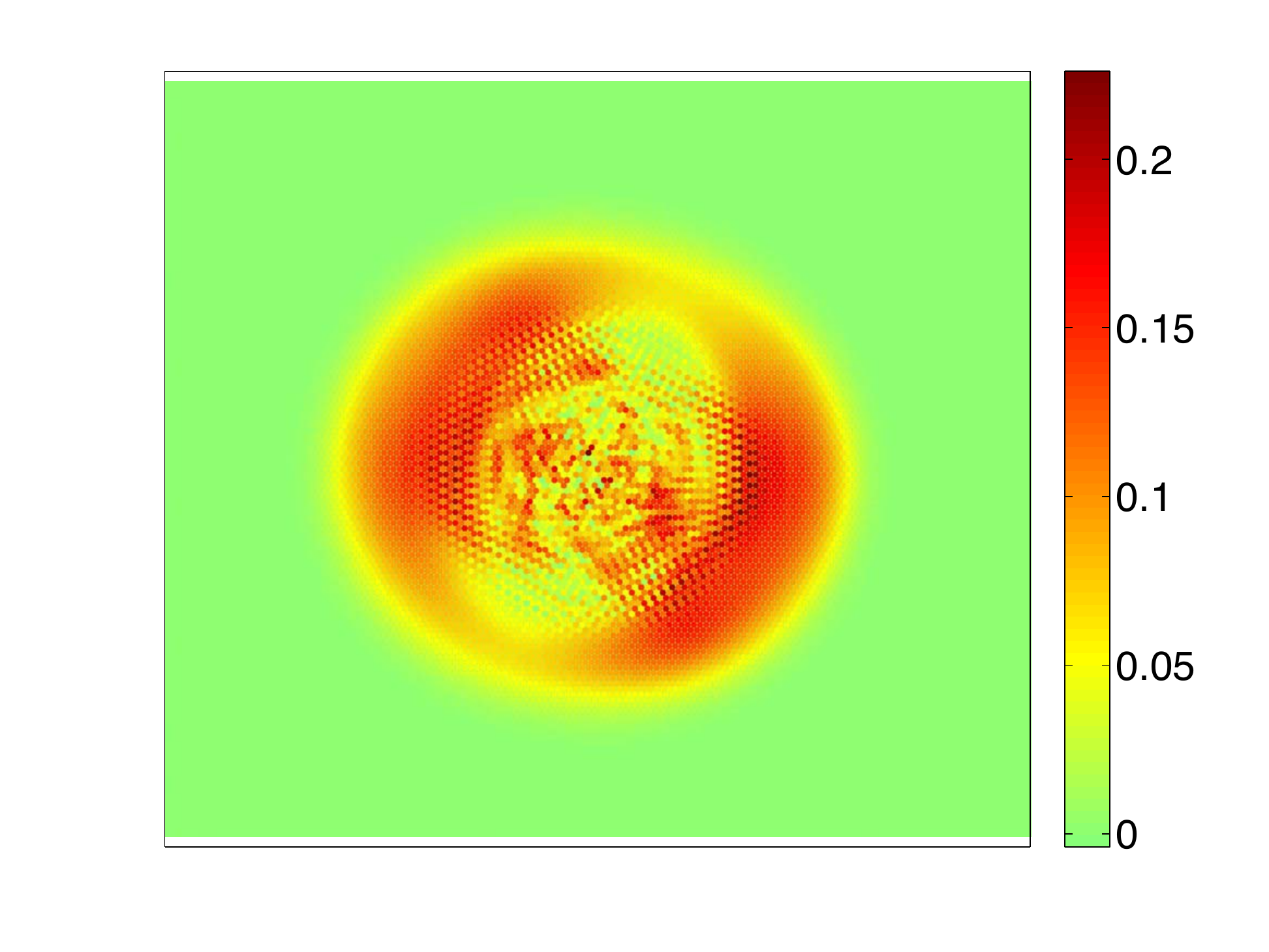} &
    \subfigimg[width=\linewidth]{\bf (i)}{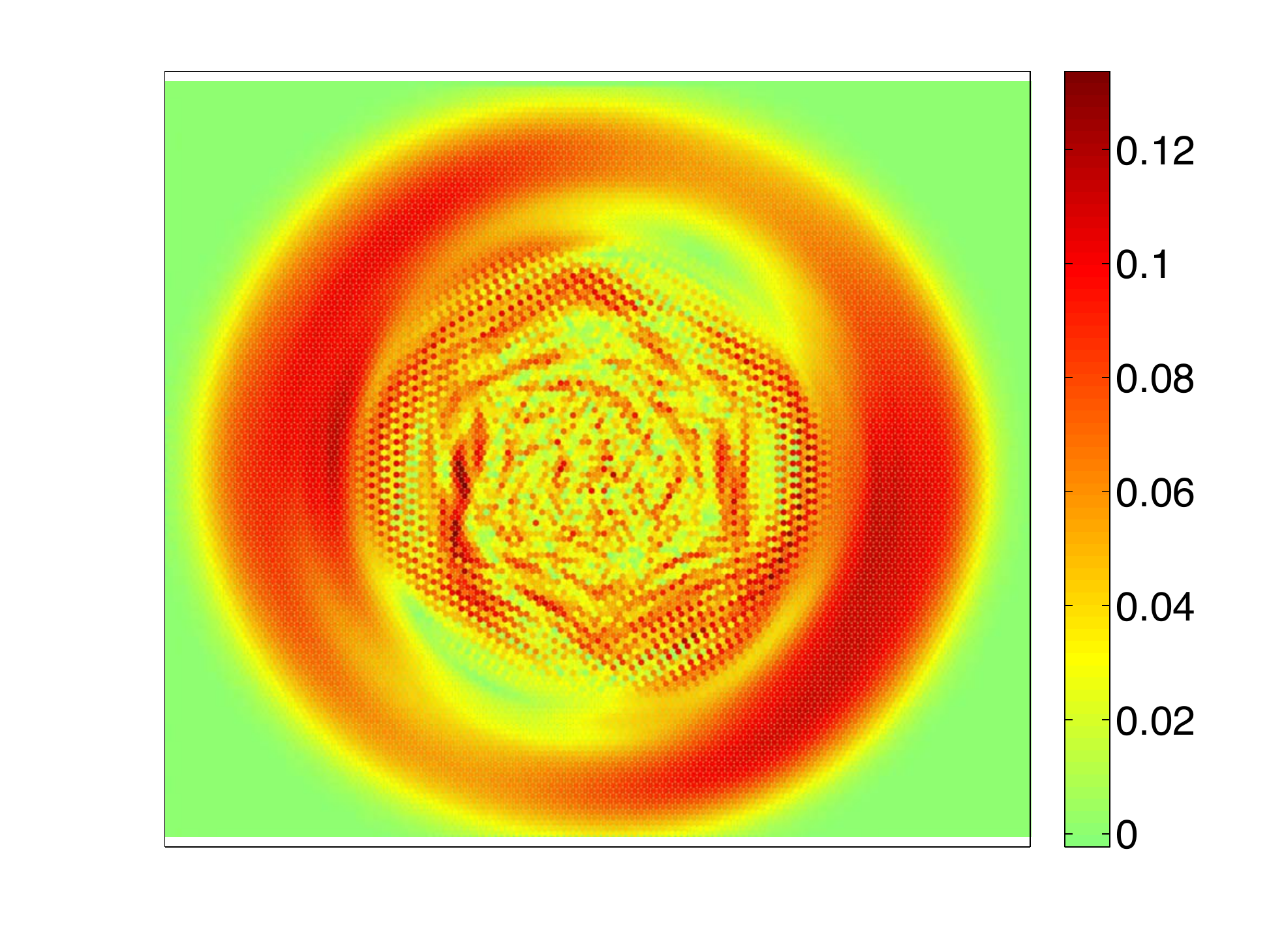}
  \end{tabular}

 \caption{Conical-like diffraction in a hexagonal lattice.   Images from left to right show the dynamical
evolution.  The color intensity corresponds to the magnitude of the
displacement. Parameters values are the same as those in Fig.~\ref{fig:spectrum}.
The initial conditions are a superposition of Bloch modes at the Dirac points $(k_d,\ell_d)=(0, 4\pi/3)$  and $(k_d,\ell_d)=(2\pi\sqrt{3}, 2 \pi/3)$,
see Eq.~\eqref{eq:IC}. Images from top to bottom correspond to increasing initial intensity.
\textbf{(a) - (c)}: Evolution with a small amplitude excitation (maximum
strain is $0.6\%$ of the static overlap $\delta$), thus inducing near linear
dynamics.
  \textbf{(d) - (f)}: Weakly nonlinear evolution (maximum strain is $10\%$ of
the static overlap $\delta$).
    \textbf{(g) - (i)}: Strongly nonlinear evolution (maximum strain is
$53\%$ of the static overlap $\delta$) .
 }
 \label{fig:cd}
\end{figure}

\begin{figure}
     \centering
   \begin{tabular}{@{}p{0.33\linewidth}@{\quad}p{0.33\linewidth}@{\quad}p{0.33\linewidth}@{}}
    \subfigimg[width=\linewidth]{\bf (a)}{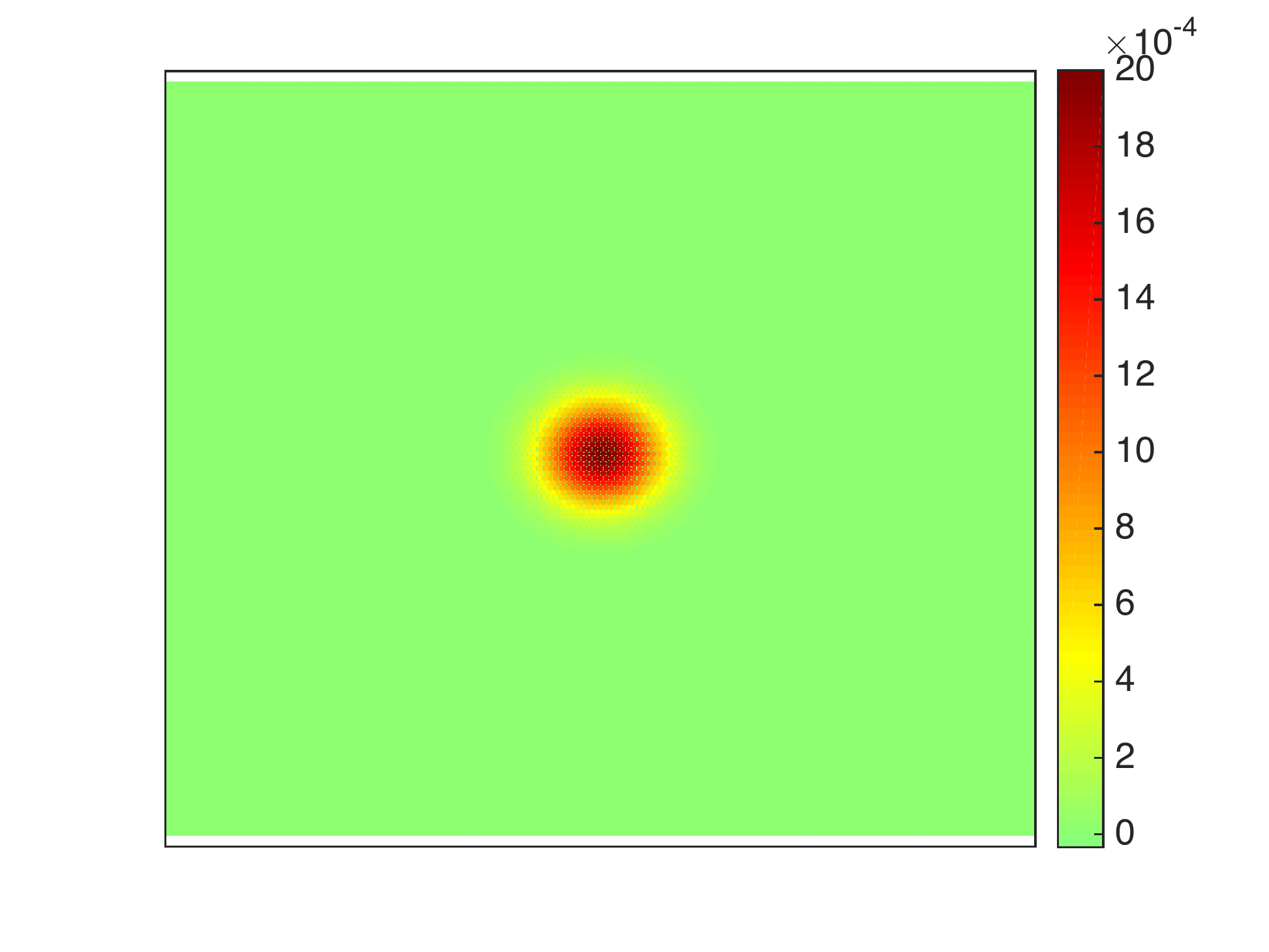} &

    \subfigimg[width=\linewidth]{\bf (b)}{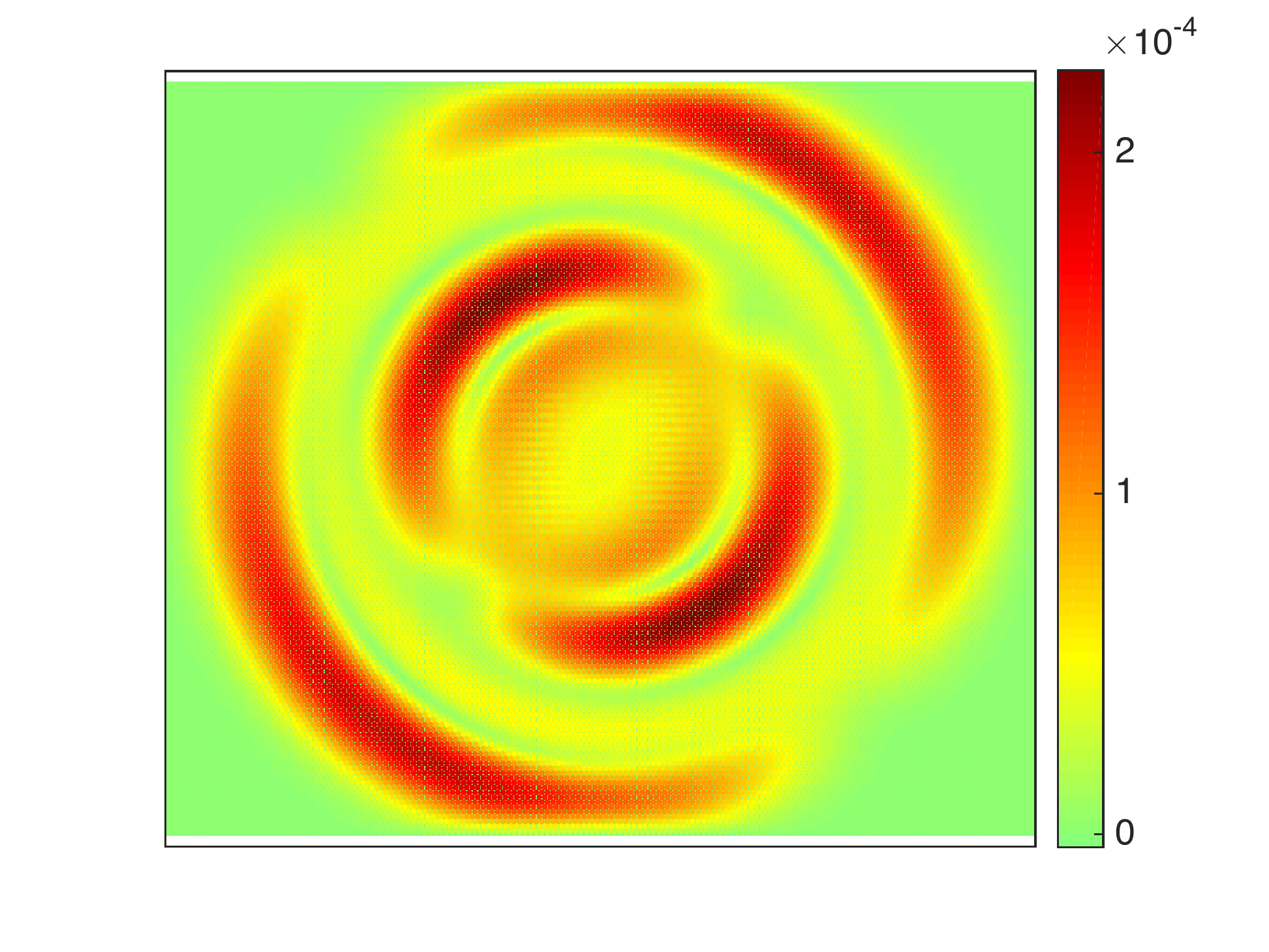} &
    \subfigimg[width=\linewidth]{\bf (c)}{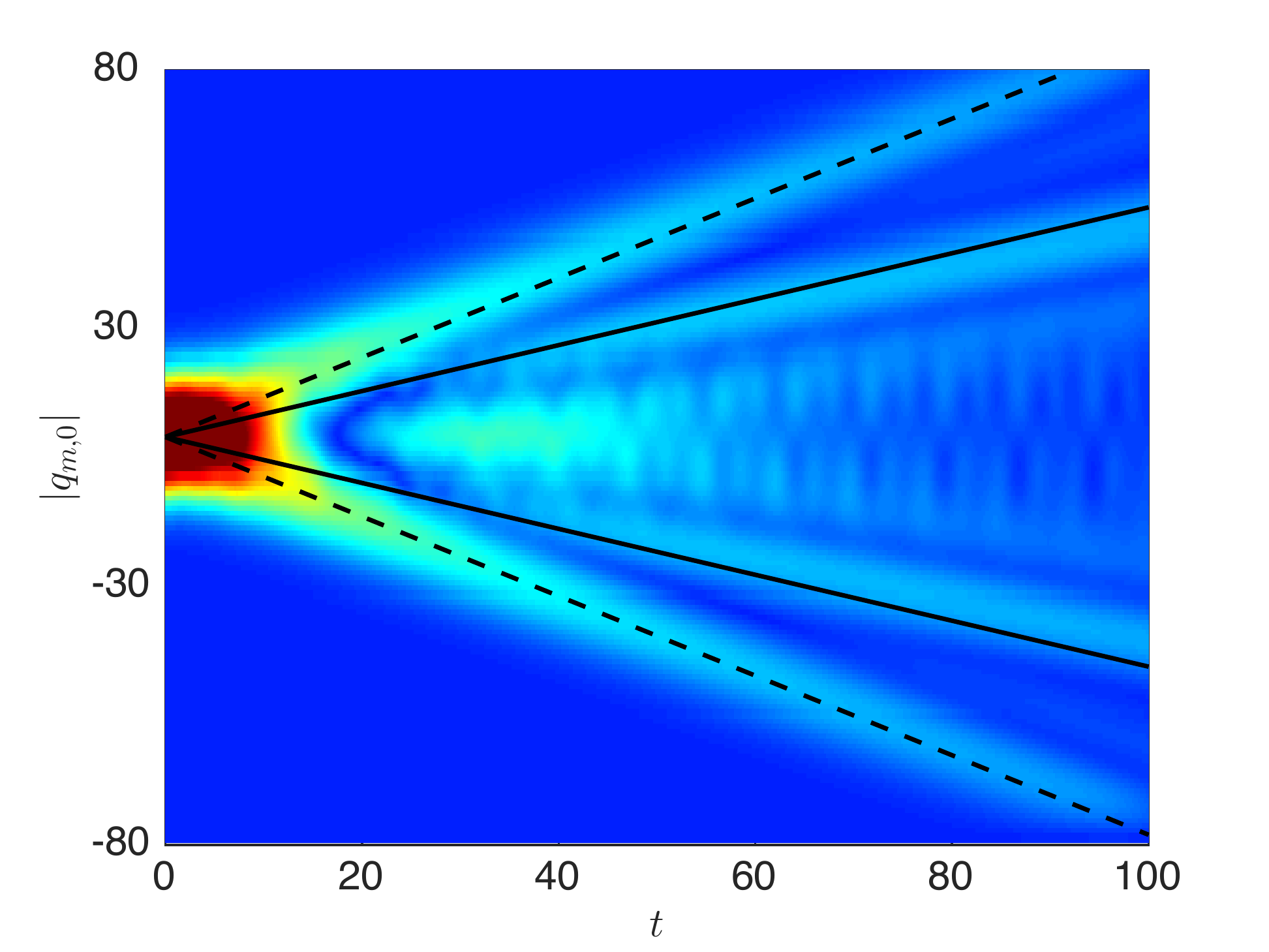}
  \end{tabular}
   \caption{ Exciting the quadruple Dirac point  $(k_0,\ell_0)=(0,0)$.   Parameters values are the same as those in Fig.~\ref{fig:spectrum}.  Note the stability condition $\delta/d < 1/3$ is satisfied.   The color intensity corresponds to the magnitude of the
displacement.
   \textbf{(a)} The initial condition with a small amplitude (i.e. near linear) excitation (maximum
strain is $0.6\%$ of the static overlap $\delta$). \textbf{(b)}  Given sufficient evolution time, two rings propagating a different speeds will form.   \textbf{(c)} Surface map of the magnitude of the displacement along the $n=0$ axis. The solid and dashed
lines are the approximations of the group speeds $\alpha_-$ and $\alpha_+$ respectively.
 }
 \label{fig:cd_double}
\end{figure}

\begin{figure}
     \centering
   \begin{tabular}{@{}p{0.33\linewidth}@{\quad}p{0.33\linewidth}@{\quad}p{0.33\linewidth}@{}}
        \subfigimg[width=\linewidth]{\bf (a)}{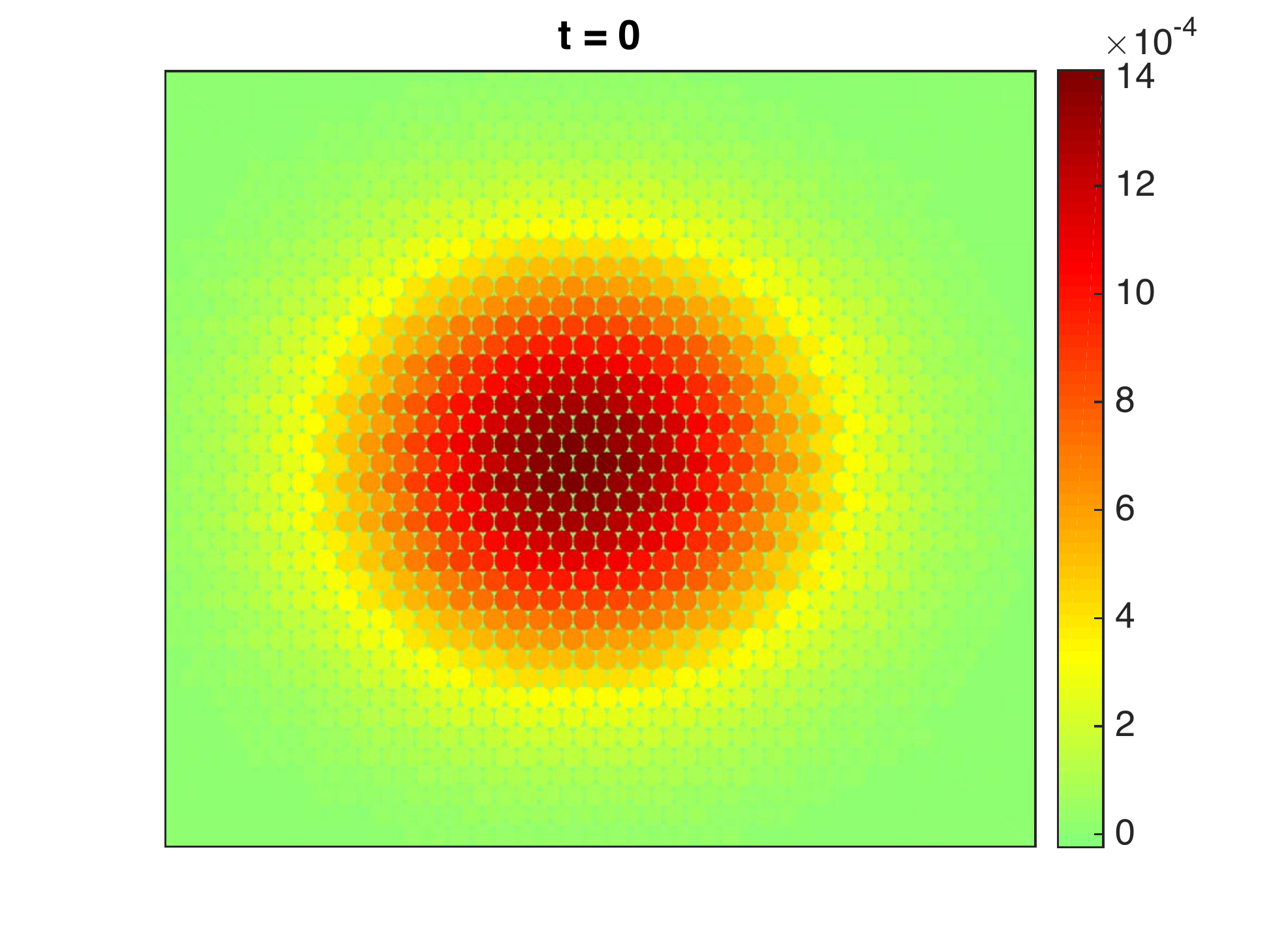} &
    \subfigimg[width=\linewidth]{\bf (b)}{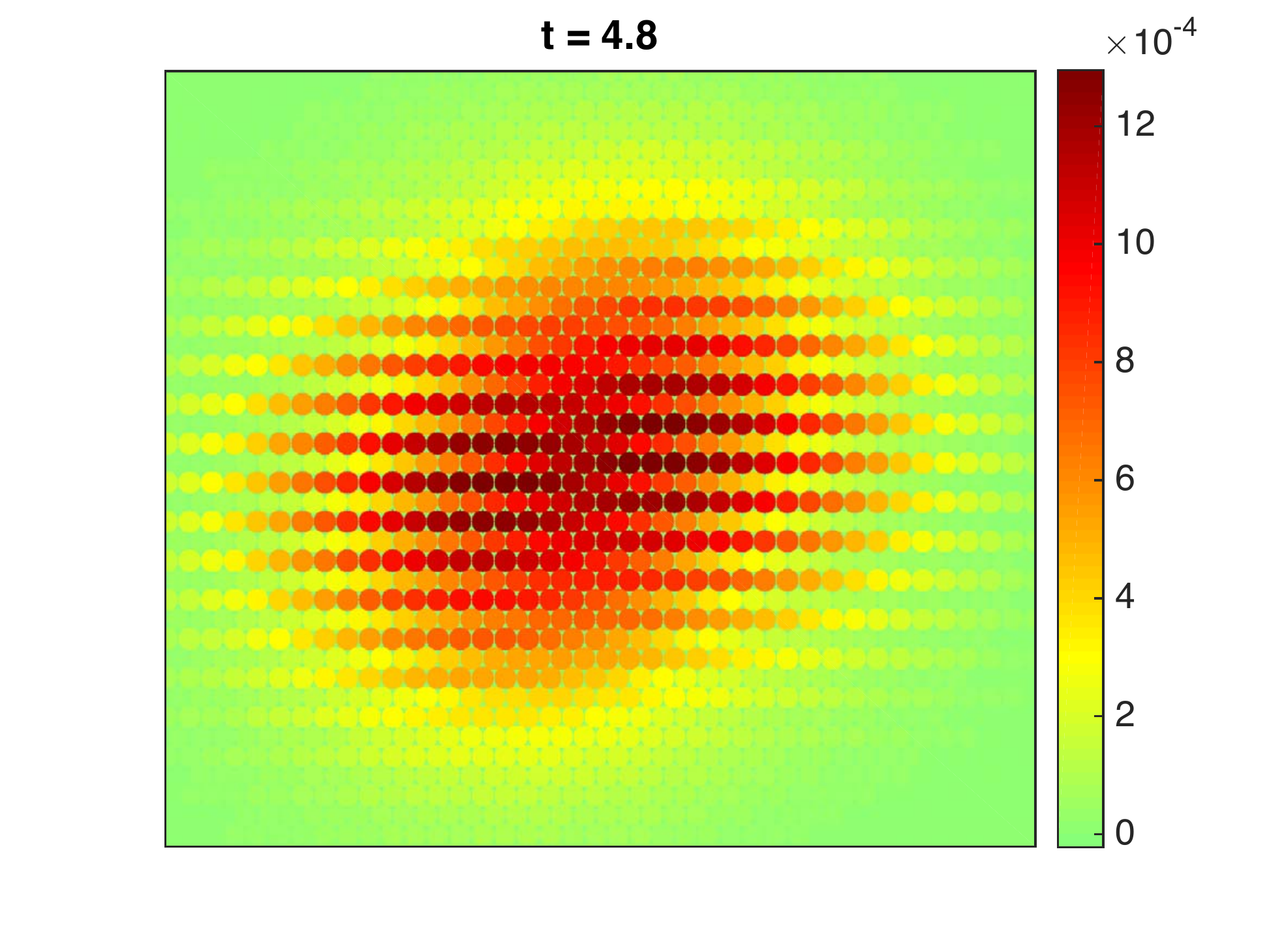} &
    \subfigimg[width=\linewidth]{\bf (c)}{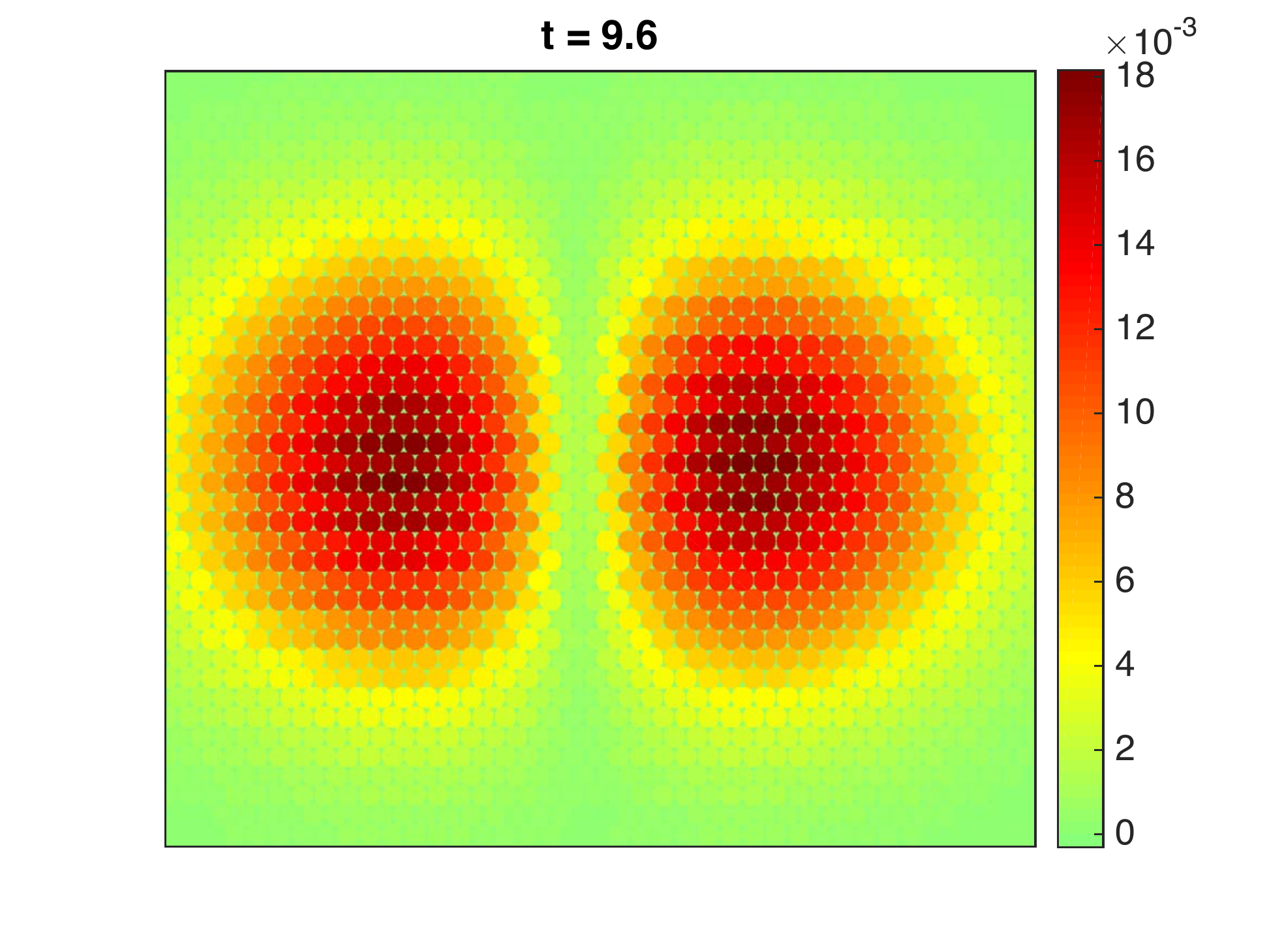}
  \end{tabular}
   \caption{  { Exciting the quadruple Dirac point  $(k_0,\ell_0)=(0,0)$ with parameters as in Fig.~\ref{fig:spectrum} with the exception of the
   precompression, which has the value $\delta = 0.4$.  The stability condition $\delta/d<1/3$ is violated (i.e. $\alpha_-$ is purely imaginary).
    \textbf{(a)} The initial condition with a small amplitude (i.e. near linear) excitation (maximum
strain is $0.6\%$ of the static overlap $\delta$). Color intensity corresponds to the magnitude of the
displacement. \textbf{(b-c)} Amplitude grows exponentially since the group speed $\alpha_-$ is purely imaginary.
The condition for validity of the model   
(which for these parameters values is $(\sqrt{3}(d-\delta) - d)/2 \approx 0.02$)
 is exceeded at $t \approx 9.7$. } }
 \label{fig:unstable}
\end{figure}

To test the conclusion of the previous sections, namely that conical diffraction is possible in discrete granular systems,
such as the hexagonally packed granular lattice,
we perform numerical simulations on a 156 x 156 packing of beads with static overlap $\delta = 0.1$. All other parameters are set to unity.
The initial condition is a localized superposition of Bloch modes near the Dirac points.  More specifically, we use
\begin{subequations} \label{eq:IC}
\begin{align}
 \begin{pmatrix} x_{m,n}(0)  \\ y_{m,n}(0) \end{pmatrix} &=  \left( \xi_1 \cos \left(  4 \pi m /3  + 2\pi n/3     \right)  \tilde{v}_1 +   \xi_2 \cos \left(  2 \pi m /3 + 4\pi n /3    \right) \tilde{v}_2  \right) A e^{-(   n^2 + m^2 + nm  )/\beta}  \\
 \begin{pmatrix} \dot{x}_{m,n}(0)  \\ \dot{y}_{m,n}(0) \end{pmatrix} &=   \left(\xi_1 \tilde{\omega}_1 \cos \left(  4 \pi m /3  + 2\pi n/3     \right)  \tilde{v}_1 +   \xi_2 \tilde{\omega}_2\cos \left(  2 \pi m /3  + 4\pi n /3    \right) \tilde{v}_2  \right)A e^{-(   n^2 + m^2 + nm  )/\beta}
 \end{align}
 \end{subequations}
where $\tilde{\omega}_1 := \omega(0, 4\pi/3)$ which has associated eigenvector $\tilde{v}_1 := v(0, 4\pi/3)$ and $\tilde{\omega}_2 := \omega(2\pi\sqrt{3}, 2 \pi/3)$
which has associated eigenvector $\tilde{v}_2 := v(2\pi\sqrt{3}, 2 \pi/3)$. $A$ is an amplitude parameter and $\beta = 100$ is the width parameter.
For $\xi_1=\xi_2=1$ the Dirac points $(0, 4\pi/3)$ and $(2\pi\sqrt{3}, 2 \pi/3)$ are excited, see Fig.~\ref{fig:cd}. The case where the single
Dirac point $(k_d,\ell_d)=(0, 4 \pi /3)$ is excited ($\xi_1=1,\xi_2=0$)  or the Dirac point $(k_d,\ell_d)=(2 \pi \sqrt{3}, 2 \pi /3)$ is excited ($\xi_1=0,\xi_2=1$)
yields qualitatively similar results and thus is not presented here.
%
To induce a linear response,
we pick $A=0.001$. In this case, as predicted by the linear theory, a ring forms, and expands throughout the lattice, maintaining its width, see Fig.~\ref{fig:cd}(a-c).
Behind this bright ring, there exists a dark ring which is also
predicted within the realm of
conical diffraction theory and is termed
the Poggendorff's dark ring~\cite{Ablowitz1,barjef}. After the dark ring,
there exists a second weaker (inner) bright ring, again as expected from the
theory of conical diffraction.
Notice also, there is a very faint (small amplitude) larger ring that expands outward in the form of a non-oscillating swell. To investigate
if the conical diffraction is hindered by the nonlinearity, we now
increase the amplitude of the excitation to $A = 0.015$. In this case,
we observe a similar set of ring structures, however,
the outer swell is of larger amplitude, see Fig.~\ref{fig:cd}(d-f). The uniform nature of the thick ring is also somewhat
altered. In the linear case, the amplitude of the thick ring is (radially)
constant
(at a fixed time), however in the weakly nonlinear case, the amplitude varies slightly. Finally, we induce a strongly nonlinear response by picking $A = 0.08$.
In this case
the outer swell is of the largest amplitude, and the inner, oscillating ring
structure is destroyed, as is the Poggendorff's dark ring.
Although we do not elaborate on the latter features further (as it is
out of scope of the present work setting the illustrative example
of the prototypical possibility of the granular lattice to sustain
conical diffraction), we do note that understanding the role of
nonlinearity would be an extremely interesting topic for future studies
on this phenomenology.
We also considered the trivial wave number $(k_0,\ell_0)=(0,0)$, where there are four Dirac cones meeting at the origin, see Fig.~\ref{fig:slice}(b). This results in
two propagating rings, each propagating with a different speed, see Fig.~\ref{fig:cd_double}(a-b). The derived
approximations of the group speeds $\alpha_-=3\sqrt{\hatd}/4$ and $\alpha_+=\sqrt{3}\alpha_-$ based on the multiple scale analysis
predicts quite well the numerically observed group speeds, see Fig.~\ref{fig:cd_double}(c). If the stability condition $\delta/d<1/3$
is violated, the excitation amplitude grows exponentially and the model's validity condition \eqref{eq:suff-no-bad-contact} is quickly exceeded, see Fig.~\ref{fig:unstable}.

\begin{figure}
     \centering
   \begin{tabular}{@{}p{0.33\linewidth}@{\quad}p{0.33\linewidth}@{}}
    \subfigimg[width=\linewidth]{\bf (a)}{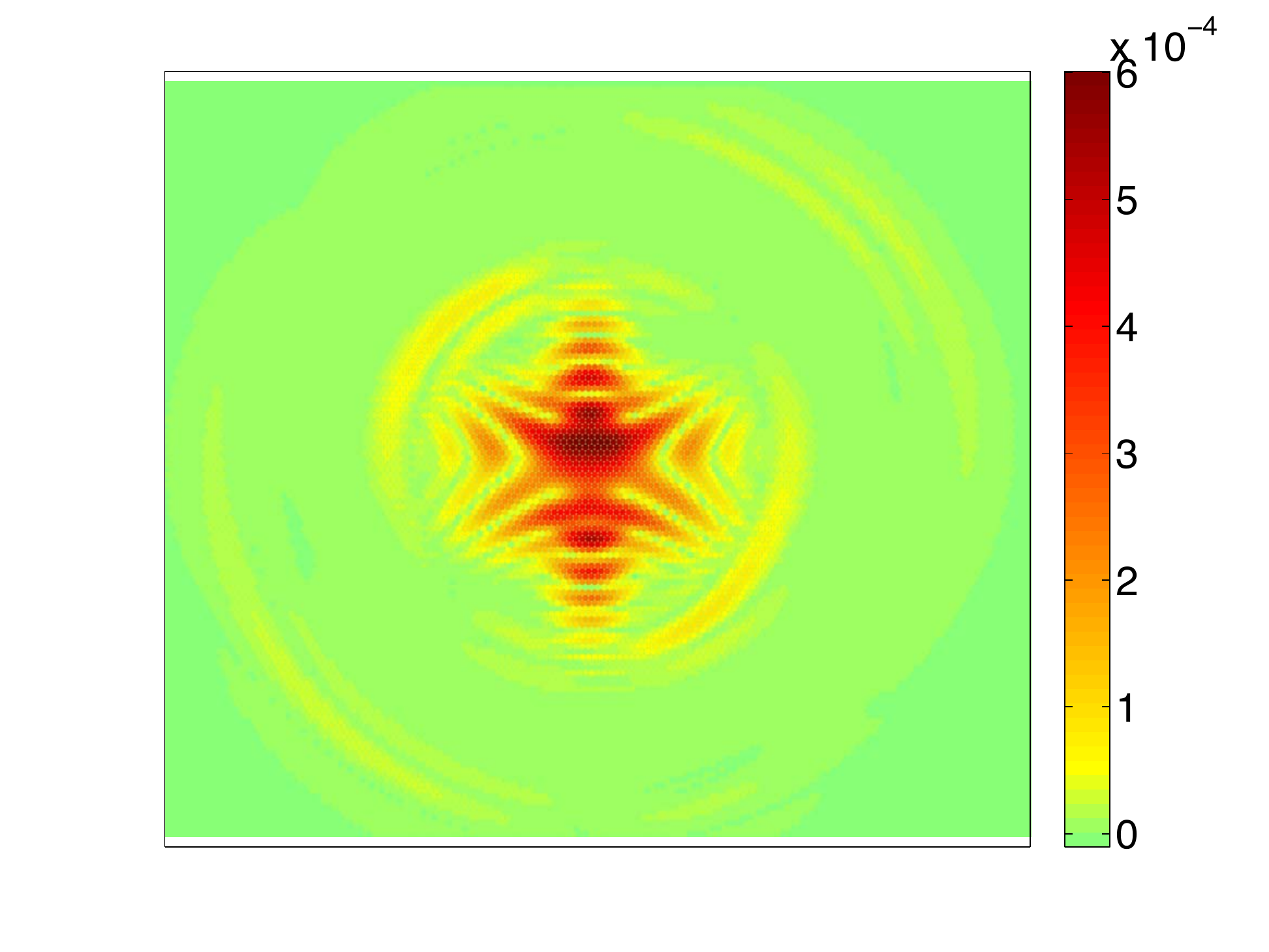} &
     \subfigimg[width=\linewidth]{\bf (b)}{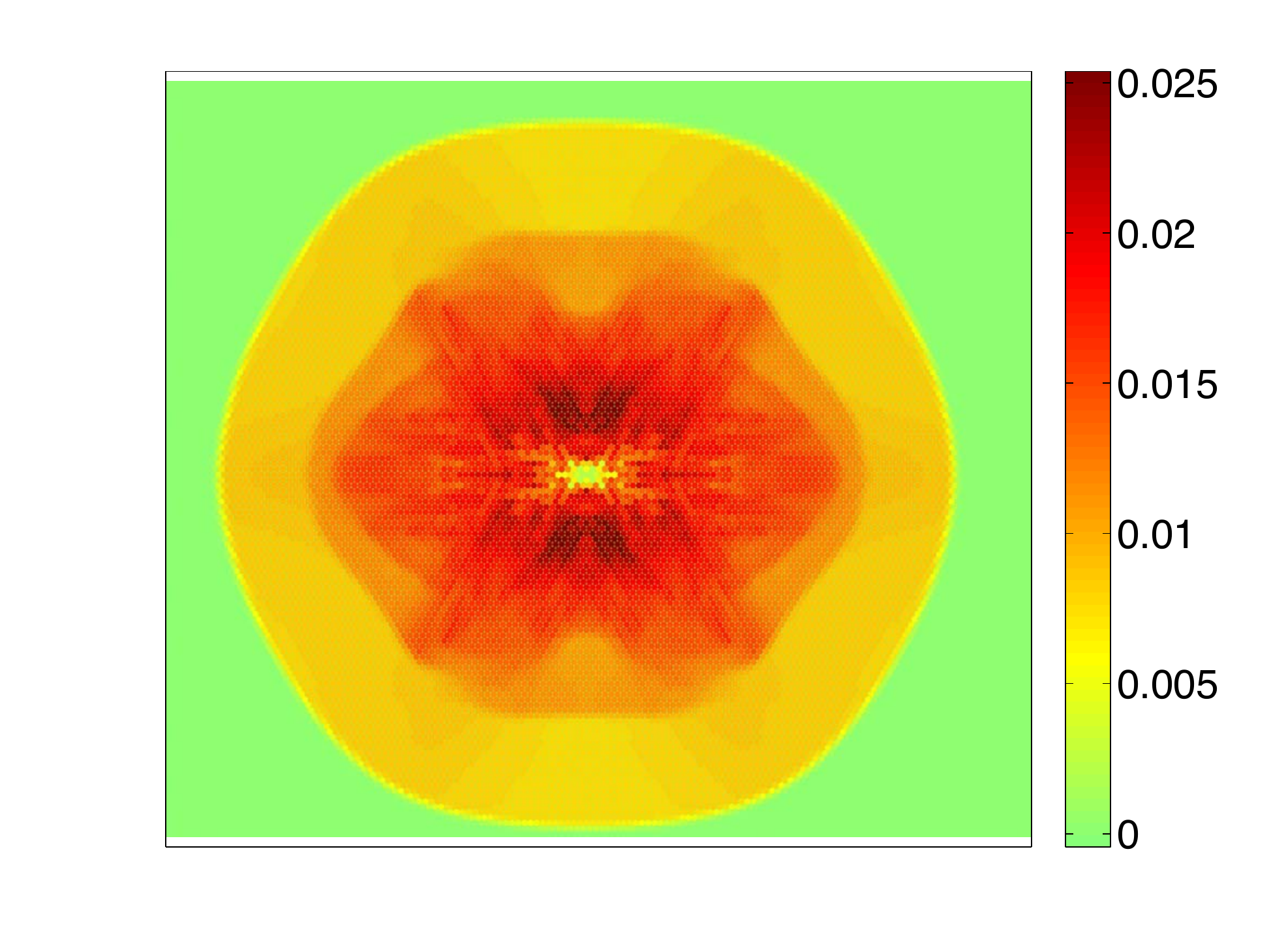}
  \end{tabular}
 \caption{\textbf{(a)} Evolution of a low amplitude Gaussian modulating a Bloch function at an arbitrary wave number.   The characteristic features
of conical diffraction are absent, while hyperbolic structures
can be discerned.  \textbf{(b)}
 Evolution after exciting four particles at the center of a purely
nonlinear chain. }
 \label{fig:arb}
\end{figure}

For the sake of comparison, we also simulate an initial condition for a Bloch wave at the arbitrary wave number $(k,\ell) = (-4,0)$ at low amplitude, see Fig.~\ref{fig:arb}(a).
Near this wave number, the local dispersion surface takes a saddle shape. In this case, conical diffraction
is not observed, but hyperbolic structures can be seen to develop.
In addition, we performed the simulation in the purely nonlinear case $(\delta = 0)$ by exciting the four beads at the center. Namely,
we considered an initial condition where all entries are zero with the exception of $\dot{x}_{0,0} = \dot{y}_{1,-1} = -0.2$ and $\dot{x}_{0,1} = \dot{y}_{0,1} = -0.2$.
In this case,
the wave front is hexagonal, but as the front becomes larger, the shape
becomes gradually more circular. Nevertheless, none of the conical
diffraction characteristic features are observed.

\section{Conclusions \& Future Challenges} \label{sec:theend}

In summary, in the present work, we provided a prototypical formulation
of the precompressed problem of granular crystals in a hexagonal configuration.
By examining the linear and weakly nonlinear regimes (as well as briefly
also venturing into the more strongly nonlinear one), we obtained in
an analytical form the linear spectrum, illustrated the existence of
Dirac points, and explored the possibility of conical diffraction
in their vicinity. We found that in the vicinity of these points
and indeed in the vicinity of the linear limit, the principal characteristics
of conical diffraction can be both derived theoretically (through
heuristic analytics, as well as through more systematic multiple
scales analysis) and observed numerically. As nonlinearity becomes
gradually more important (or as we depart from these points),
this phenomenology gets progressively modified and eventually it
appears to break down in the presence of most substantial nonlinear
interactions.

Naturally, a considerable volume of possibilities emerges
from this initial study. Perhaps one of the most interesting
aspects is to explore in further detail both the weakly
and the strongly nonlinear regime. In the former, although
technically rather cumbersome, it would seem to be very worthwhile
to explore the nonlinear version of the Dirac equation that
a multiple scales analysis should produce. From an experimental
perspective, these systems appear to be well within reach
since either in the realm of beads~\cite{Leonard11,andrea},
or even in the more recent setup of magnets~\cite{miguel}, it should
be possible to
construct a system tantamount to the one considered here, bearing
in mind the considerable insights that their optical (even linear)
analogues
have offered; for a recent example, see~\cite{zhigang}.
Finally, this realization, in turn, would pave the way for
additional intriguing features such as potential
acoustic realizations~\cite{topac} of topological edge states~cf. \cite{rechts13,acm14,mjaypm,Huber15},
among others. These themes are currently under study and will be reported
in future publications.

\section*{Acknowledgements}
 C.C. was partially supported by the ETH Zurich Foundation through the Seed Project ESC-A 06-14. The research of M.J.A. and Y-P Ma was partially supported by NSF under Grant No. DMS-1310200.
The work of P.G.K. at Los Alamos is partially supported by
the US Department of Energy. P.G.K. also gratefully acknowledges
the support of BSF-2010239, as
well as from the US-AFOSR under grant FA9550-12-1- 0332,
and the ERC under FP7, Marie Curie Actions, People, International
Research Staff Exchange Scheme (IRSES-605096).

\end{document}